\documentclass[12pt,preprint]{aastex}
\shorttitle{INFRARED CLOUD MONITOR FOR ROBOTIC TELESCOPE}
\shortauthors{Suganuma et al.}

\begin{document}

\title{THE INFRARED CLOUD MONITOR FOR THE MAGNUM ROBOTIC TELESCOPE AT HALEAKALA}

\author{Masahiro Suganuma\altaffilmark{1,2},
    Yukiyasu Kobayashi\altaffilmark{1},
    Norio Okada\altaffilmark{1},
    Yuzuru Yoshii\altaffilmark{3,4}, 
    Takeo Minezaki\altaffilmark{3}, 
    Tsutomu Aoki\altaffilmark{5}, 
    Keigo Enya\altaffilmark{6},
    Hiroyuki Tomita\altaffilmark{5}, 
    and
    Shintaro Koshida\altaffilmark{1,7}}

\altaffiltext{1}{National Astronomical Observatory,
    2-21-1 Osawa, Mitaka, Tokyo 181-8588, Japan}

\altaffiltext{2}{e-mail: suganuma@merope.mtk.nao.ac.jp}

\altaffiltext{3}{Institute of Astronomy, School of Science, University of Tokyo,
    2-21-1 Osawa, Mitaka, Tokyo 181-0015, Japan}

\altaffiltext{4}{Research Center for the Early Universe,
    School of Science, University of Tokyo,
    7-3-1 Hongo, Bunkyo-ku, Tokyo 113-0033, Japan}

\altaffiltext{5}{Kiso Observatory,
    Institute of Astronomy, School of Science, University of Tokyo,
    10762-30 Mitake, Kiso, Nagano 397-0101, Japan}

\altaffiltext{6}{Institute of Space and Astronomical Science, 
    Japan Aerospace Exploration Agency, 
    3-1-1, Yoshinodai, Sagamihara, Kanagawa, 229-8510, Japan}

\altaffiltext{7}{Department of Astronomy, School of Science, University of Tokyo,
    7-3-1 Hongo, Bunkyo-ku, Tokyo 113-0013, Japan}

\begin{abstract}

We present the most successful infrared cloud monitor for a robotic telescope.
This system was originally developed for the MAGNUM 2-m telescope, which has been achieving unmanned and automated monitoring observation of active galactic nuclei at Haleakala on the Hawaiian island of Maui since 2001.
Using a thermal imager and two aspherical mirrors, it at once sees almost the whole sky at a wavelength of $\lambda\sim 10\mu{\rm m}$.
Its outdoor part is weather-proof and is totally maintenance-free. 
The images obtained every one or two minutes are analysed immediately into several ranks of weather condition, from which our automated observing system not only decides to open or close the dome,
but also selects what types of observations should be done.
The whole-sky data accumulated over four years show that 50$-$60 \% of all nights are photometric, and about 75 \% are observable with respect to cloud condition at Haleakala.
Many copies of this system are now used all over the world such as Mauna Kea in Hawaii, Atacama in Chile, and Okayama and Kiso in Japan. 

\end{abstract}



\section{INTRODUCTION}


A cloud monitoring system, which watches the sky to detect clouds above an observatory, is a powerful apparatus for ground-based telescopes if we want to check the sky easily and to execute some remote or automated observations.
The telescopes and their instruments are sure to be safe if we can monitor clouds on time and can close the dome slit before cloud coverage becomes heavy and rain drops come.
Other types of weather sensors, such as rain or humidity sensors, are sometimes late to alert us to close the dome.
With a cloud monitor, we also can always be sure whether the data acquired by the telescope has been affected by clouds or not.

One of the smartest methods of seeing clouds from the ground is to use some thermal infrared wavebands in which clouds themselves emit thermal radiation or reflect radiation originating from ground or sea.
A CCD camera with a fish-eye lens is cheap, but the appearance of clouds in optical is deceptive because their brightness depends strongly on the intensity of the moon and city lights that illuminate them.
Using a similar system, Shamir \& Nemiroff (2005) developed an algorithm to make a whole-sky opacity map by means of measuring the extinctions for many stars.
However, it does not give us a direct view of the cloud distribution in the sky.

An uncooled thermal imager with panoramic optics suits a robotic telescope because it requires little maintenance.
The Sloan Digital Sky Survey (SDSS) project has developed a scanning system using a single channel photometer cooled by liquid nitrogen (Hull et al. 1994; Hogg et al. 2001).
It has enough sensitivity and field of view, but their scanning mechanism is not easy to construct and its cooling system needs frequent hands.
Recently, thermal infrared imagers without cooling parts sensitive enough to detect thin clouds have become available, and by combining these with some panoramic optics, now we can easily see the whole sky in thermal infrared. 
This idea was presented by Mallama \& Deganan (2002), and a similar one was developed by the Apache Point Observatory\footnote{
http://irsc.apo.nmsu.edu/
}.
However, it requires stability and reliability for these systems to be put into practical use for a robotic telescope where no operator or engineer is onsite.

We developed an infrared cloud monitor system that is successful for this use. 
Owing to this, we have achieved unmanned automated observation at the MAGNUM observatory at Haleakala since 2001 (Kobayashi et al. 2003; 2004).
The MAGNUM (\underline{M}ulticolor \underline{A}ctive \underline{G}alactic \underline{Nu}clei \underline{M}onitoring) project newly built a 2-m optical-infrared telescope at the University of Hawaii's Haleakala Observatory site on the Hawaiian Island of Maui, and has been monitoring many active galaxies and quasars in optical and near-infrared wavebands for more than several years (Yoshii et al. 2003). 
We study their structure and physical environment, and finally, determine the cosmological parameters by an entirely original method (Kobayashi et al. 1998; Yoshii 2002).
The other distinct challenge of this project is to achieve unmanned automated observation.
We aimed at and have achieved months-long automated observation without anyone being required at the observatory. 

In order to fulfill our purpose, our whole-sky infrared cloud monitor has many salient characteristics for automated operation.
It can see almost whole-sky at once in a thermal infrared waveband with sensitivity high enough to detect thin clouds.
It is weather-proof, and has been outdoors under hard weather conditions on the top of a high-altitude mountain, about 3,000m high, for many years.
Furthermore, the raw whole-sky images are immediately reduced to apparent emissivity maps of the cloud which will be classified into several ranks of observational condition.
These maps and ranks can be referred to by our automated observing system and the remote watchers in Japan.
They are also referred to by many facilities other than the MAGNUM telescope at Haleakala.
Following our success, systems using copies of our design have been used at several observatories and sites such as Mauna Kea in Hawaii (Takato et al. 2002), Atacama in Chile, and Okayama and Kiso in Japan. 
These attract site studies for new observatories where it is difficult for people to remain for any length of time.

In this paper, we describe our infrared cloud monitor system, breaking down into its design, its hardware contents, its data analysis software, and its performance and statistical data. 
The main instrument design and its components including the thermal imager, reflecting optics, and data acquisition system are described in $\S$ \ref{system}. 
In $\S$ \ref{software}, the analysis software that detects clouds and evaluates the whole-sky condition is described.
The performance in some respects in operating the MAGNUM observatory is presented and discussed in $\S$ \ref{performance}. 
Finally, the weather trends seen in our accumulated whole-sky condition data over four years are discussed in $\S$ \ref{weather_trends}.
\\


\section{System Overview}
\label{system}

\begin{figure}[htbp]
  \epsscale{1.0}
  \plotone{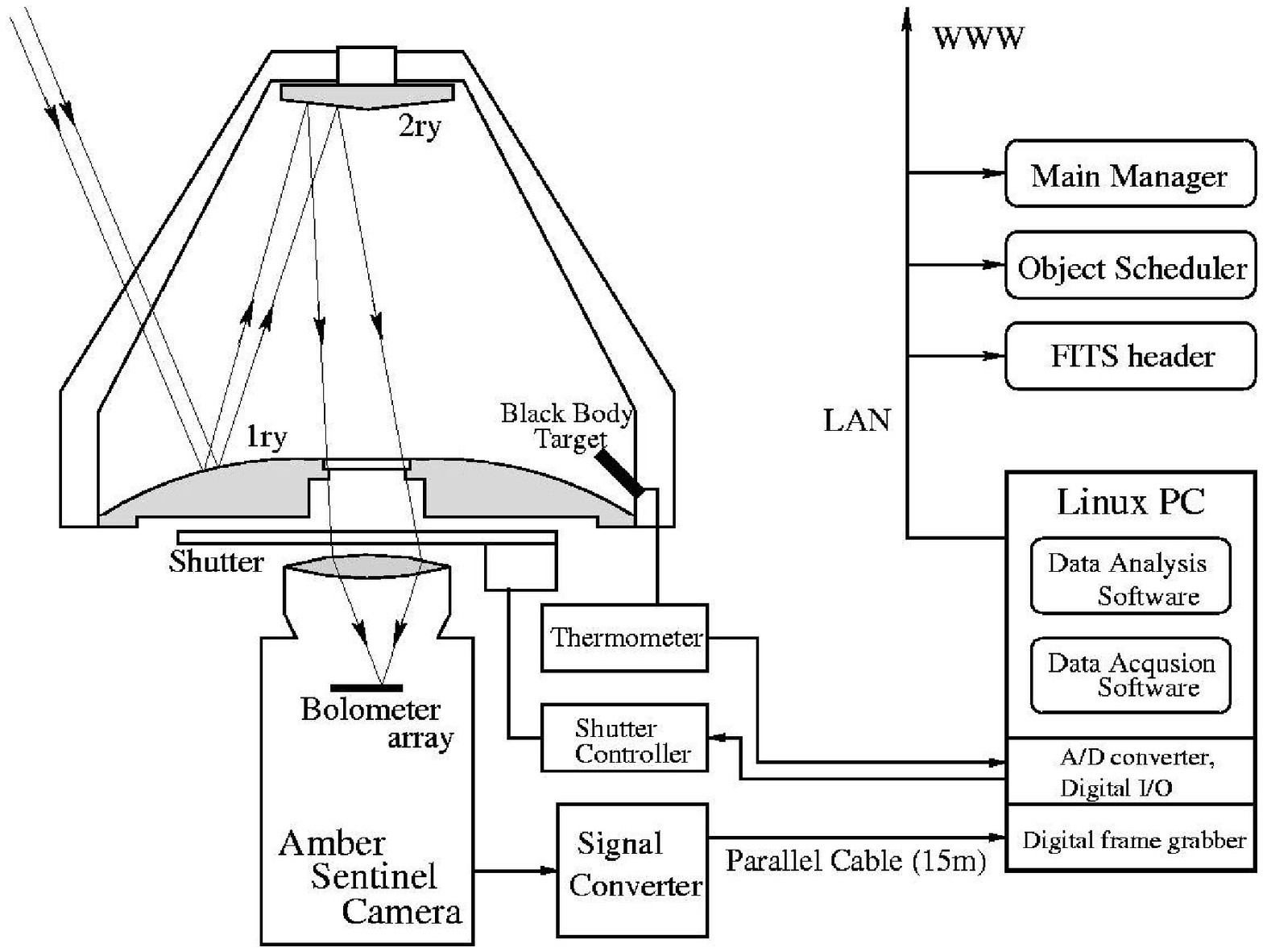}
  \caption{Schematic diagram of MAGNUM infrared cloud monitor system; The outdoor hardware is to the left of the figure and the indoor hardware is to the lower right.}
  \label{sysicm}
\end{figure}
\begin{figure}[htbp]
\epsscale{0.8}
  \plotone{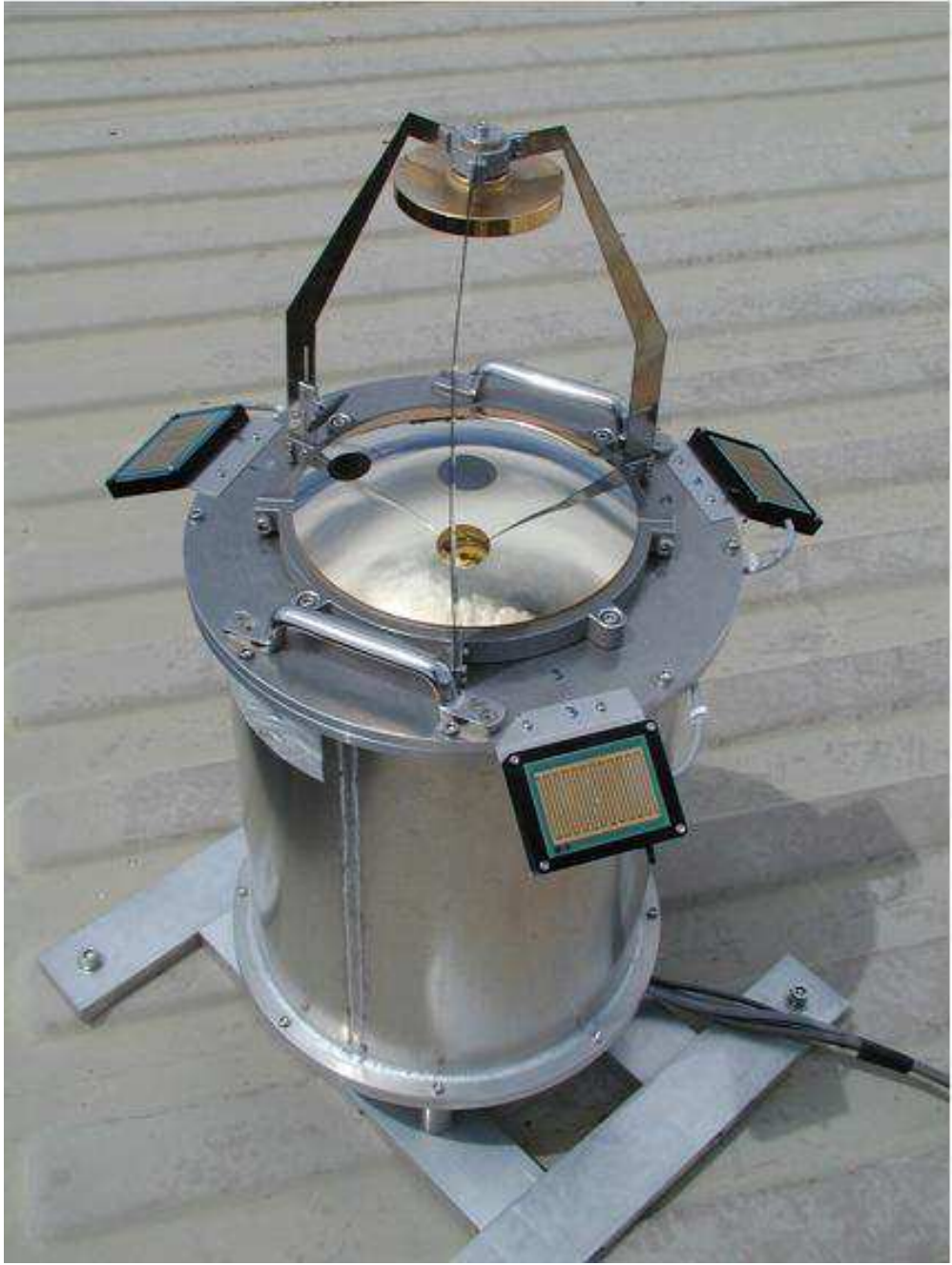}
  \caption{Outdoor part of MAGNUM infrared cloud monitor. 
   Shown from top to bottom are a secondary mirror, a primary mirror, and an aluminum-pipe housing that contains a thermal imager, a signal converter, a shutter controller, etc. 
   The three rectangle plates extending from the upper edge of the housing are rain sensors that form a different system from the cloud monitor, but share power and wires with it.}
  \label{magicm}
\end{figure}
\begin{figure}[htbp]
\epsscale{0.9}
   \plotone{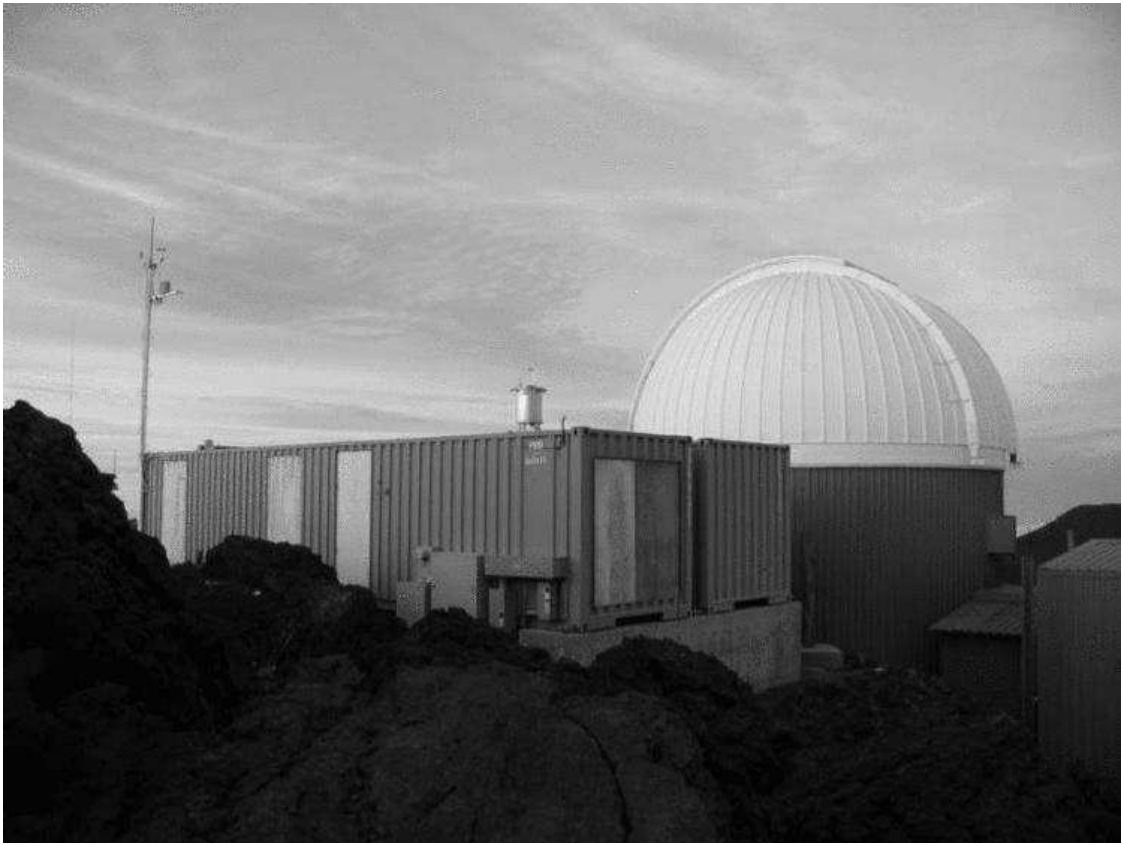}
  \caption{Infrared cloud monitor at MAGNUM Observatory. The cloud monitor is seen near the center, on the roof of the container.}
  \label{weath1}
\end{figure}

Figure \ref{sysicm} shows a schematic diagram of the MAGNUM Infrared Cloud Monitor. 
The outdoor hardware of the system is to the left of the figure.
There are two aspherical convex mirrors with Cassegrain-like alignment. 
A blackbody reference plate for calibration is installed where the camera can see it near the edge of its field of view.
All electrical devices, thermal imager, signal converter for the output signal of the imager, shutter, and thermometer circuit for the blackbody reference plate, are attached under the primary mirror and surrounded by an aluminum-pipe housing whose ceiling is the primary mirror. 
On the central hole of the primary mirror is installed a diamond window which is transparent to thermal infrared and prevents water from dripping into the housing. 
Photographs of the outdoor hardware are shown in Figures \ref{magicm} and \ref{weath1}. 
It is about 80 cm in height and about 35 cm in diameter.

The data observed by the outdoor hardware are acquired by a Linux PC, illustrated on the lower right of Figure \ref{sysicm}.
The PC controls the shutter with a digital I/O board, and also triggers data acquisition of images and temperatures. 
The controlling and data-acquiring software, as well as analysis software, are always working and getting sets of the data including whole-sky raw images every one or two minutes throughout the night.
The output data are directed to various types of software through a LAN in the MAGNUM observatory that includes the main manager of the observatory, the real-time scheduler or selector of target astronomical objects, and the information-collecting server for image headers of astronomical observations\footnote{
             GIF images of the whole-sky emissivity cloud maps can be found at http://banana.ifa.hawaii.edu/cloud/
}.

The specifications of our infrared cloud monitor are listed in Table \ref{speicm}. 
Owing to the uncooled thermal imager and reflecting optics that widen the camera field of view, we can obtain almost whole-sky images in thermal infrared.
This hardware alignment, a Cassegrain-like mirror system above an aluminum-pipe housing containing the thermal imager and electronic parts, has a significant advantage in that the outdoor system becomes compact and is waterproof.
The only movable component in the system is a shutter, the ``PRONTER magnetic E/100", which works both as a shield against the sunlight during daytime and as a flat-fielding plate for image reduction.

The main hardware components of our infrared cloud monitor are the thermal imager, the reflecting optics, and the data acquisition system, the details of which are described individually in the next subsections.
\\


\subsection{Thermal Imager}
\label{imager}


The uncooled infrared imager that is sensitive in the $10\mu {\rm m}$ waveband is one of the key components of our cloud monitor.
In order to see high-altitude clouds with good visibility, or to detect the thermal emission of clouds standing out against the dark background of cold space, we need to select a waveband in which the atmosphere is highly transparent.
There are two wavebands satisfying this requirement: one is the $3\sim 5$ $\mu $m band, and the other is the $10$ $\mu $m band.
So far as we use only one waveband and cannot measure cloud temperature, the $10$ $\mu $m band is better to estimate cloud emissivity. 
This is because this wavelength is around the flat-top part of the blackbody radiation of the clouds and ground, and flux from the clouds in this band is less dependent on temperatures than in the $3\sim 5$ $\mu $m band.

We use an Amber Sentinel Camera commercially produced by the Amber A. Raytheon Co. in 1997. 
This imager has an uncooled bolometer array of 320$\times$240 pixels for its detector, and is sensitive enough to detect thin cirrus clouds in thermal infrared.
It outputs both an analog signal of NTSC standard and digital signals of 12-bit depth with a frame rate of 30 Hz.
The specifications of the imager are tabulated in Table \ref{sentinel}.

The camera has an automatically offset flat-fielding function. 
It compensates for pix-to-pix scatter of bias and dark current signals that depend heavily on the temperature of the imager.
This calibration is necessary for a quick look at raw images and is recommended every several minutes; calibration can be triggered by PC through an RS232C interface.
However, we do this calibration less frequently, and take integrated images of a closed shutter plate for more precise compensation (see \S\ref{reduction}).

Several cautions are in order here.
Some imagers could not take clear images in a cold environment, since their detector outputs are reduced outside the operation range.
Generally, outputs of uncooled bolometers significantly change due to a large change of thermal background, because they are surrounded by internal parts of ambient temperature.
Imagers being optimized for use at room temperature have relatively narrow operation ranges, and sometimes have the problem under low temperature environment. 
Our imager displayed this defect in winter in Japan or at the Haleakala site, so we asked the company to tune the electric circuit of the imager in colder environments.
Moreover, we resistively heated the imager up to about room temperature at night to increase the thermal background signals from the imager itself.

Incidentally, the imager cost as much as \$40,000 when we developed our cloud monitor.
Recently, products with various specifications have been made available by many manufacturers at lower costs.
\\


\subsection{Reflecting Optics}
\label{optics}


\subsubsection{Design of Optics}

The Cassegrain-like reflecting optics is another key component of our infrared cloud monitor.
The camera's field of view is not large enough to cover the whole sky, and must be expanded by other optics. 
Germanium crystals are generally used for the lenses of these thermal imagers because it is well transparent in thermal infrared. 
Fish-eye lenses made of this material, however, are not commercially available, and are also hard to develop or fabricate without a great deal of expenditure.
Thus, it is reasonable to use some kind of convex mirror system.

For the fundamental shape of the mirrors, we based ours on the particular aspherical ones introduced by Chahl \& Srinivasan (1997).
When we look into these types of mirrors along with their optical axis, the appearance of the reflected field is not radially deformed; they preserve a linear relationship between the apparent angle from the image center and the real radial angle from the field center.
A simple spherical mirror produces a radially compressed image toward the image edge; the more distant from the image center we see through the mirror, the more radially compressed the objects appear.

\begin{figure}[htbp]
  \epsscale{0.6}
  \plotone{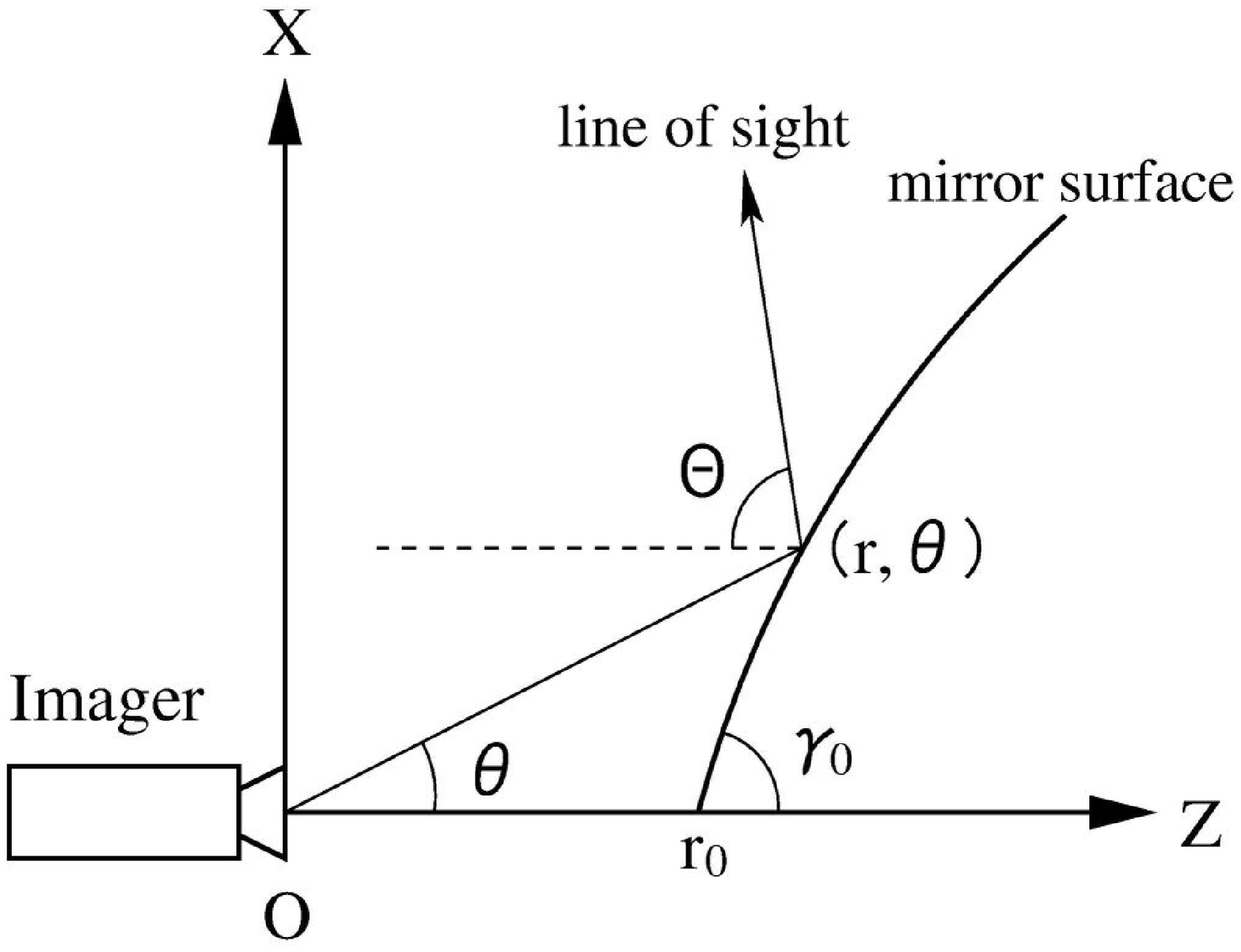}
  \caption{Basic alignment of the imager and the section of the panoramic aspherical mirror surface introduced by Chahl \& Srinivasan (1997).}
  \label{opt_design1}
\vspace{0.3in}
  \epsscale{0.75}
  \plotone{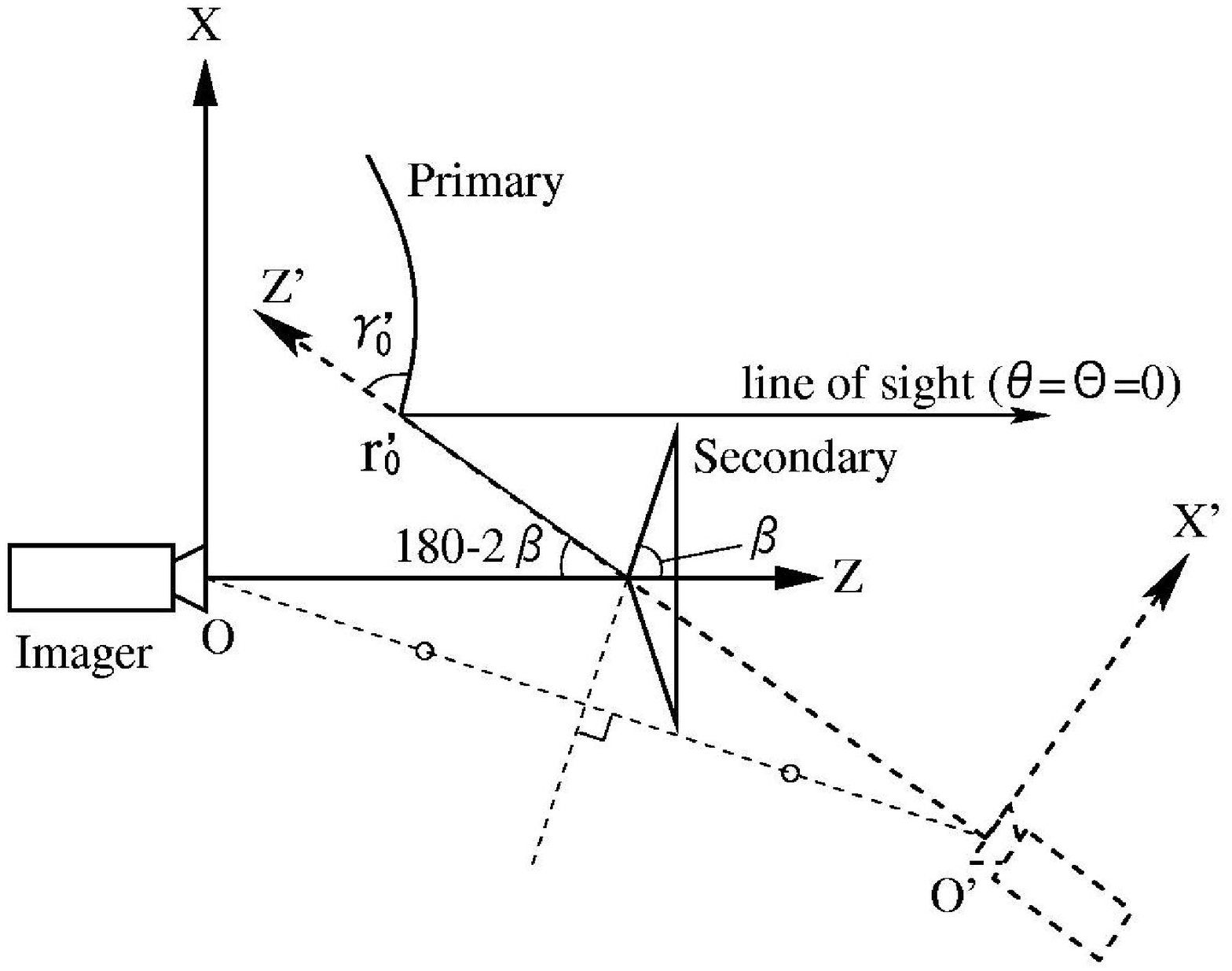}
  \caption{The basic alignment of the imager and the section of Cassegrain-like mirror system that was originally introduced.
             }
  \label{opt_design2}
\end{figure}

Now, we align the camera with the surface of a mirror using polar coordinates as shown in Figure \ref{opt_design1}, where the lens node of the imager is located at the origin, and the mirror surface should be adjusted by the revolution of the function $r(\theta)$ around $Z$ axis.
Here, $\theta$ is an angle of line-of-sight with the $Z$ axis in the camera field, and $\Theta$ is an angle of line-of-sight with the $Z$ axis in the negative direction in the real field that is seen through both the camera and the mirror.  
According to Chahl \& Srinivasan, $\Theta$ would proportionally correspond to $\theta$ if $r(\theta)$ is given by
     \begin{equation}
     r(\theta)=\frac{r_0 (\sin\gamma_0)^{-1/\kappa}}{\left[ \sin(\kappa\theta+\gamma_0) 
     \right]^{-1/\kappa} } \;\;\;,
     \end{equation}
where $r_{0}$ is the distance between the mirror and $O$ along the $Z$ axis, $\gamma_{0}=\tan^{-1}[dr(\theta=0)/dz]$ is the initial angle of the mirror, in other words, a half of the vertex angle, and $\kappa$ relates to the proportionality constant $\alpha$ between $\theta$ and $\Theta$, which means field widening power, as
     \begin{equation}
     \frac{d\Theta}{d\theta} = -1-2\kappa = \alpha\;\;\;.
     \end{equation}
If we place the mirror with the convex side upwards and direct the camera to look down on the mirror vertically, we can see the sky at zenith angle between $\Theta_{\rm min}$ and $\Theta_{\rm max}$ in degrees:
     \begin{eqnarray}
     \Theta_{\rm min}&=&2(90^{\circ}-\gamma_{0})\;\;\;\;\;\;\;\;\;\\
                           &{\rm and}&\nonumber\\ 
     \Theta_{\rm max}&=&\theta_{\rm max} + 2\left( 90^{\circ}-\tan^{-1}\left[\frac{dr(\theta=\theta_{\rm max})}{dz}\right] \right)\;\;\;.
     \end{eqnarray}
A circular image of the whole-sky is obtained by setting $\theta_{\rm max}$ as half the shorter angle of the rectangular field-of-view of the imager.

Chahl \& Srinivasan (1997) also suggested a two-mirror system of Cassegrain-like alignment, in which the aspherical shape introduced above is used as a primary mirror and a cone-like shape is used as secondary, as shown in Figure \ref{opt_design2}.
Here, the secondary with half a vertex angle of $\beta$ is placed with its surface facing the imager, which is equivalent to the configuration in Figure \ref{opt_design1} with $\gamma_{0}=\beta$.
The section of its surface is triangular, equivalent to $\alpha=1$, which means the surface has no field widening power.
The section of primary mirror surface should be drawn similarly to Figure \ref{opt_design1}, but in an $X'O'Z'$ coordinate system, in which $O'$ is symmetrical with $O$ about the section of the secondary surface.
If $\beta$, $r'_{0}$, and $\gamma'_{0}$ are properly optimized, the light coming from the zenith ($\theta=\Theta=0$) can reach the imager, avoiding the secondary mirror by way of $r'_{0}$ and the secondary vertex.
Then, the entire sky including the zenith can be seen.

However, though the idea is very attractive, we found a serious astigmatism aberration in these optics which is mostly derived from the cone-shape of the secondary.
When seen from above along with the optical axis, there is no curvature along with sagittal directions on the completely cone-shaped surface, while its curvature exists tangentially.
This aberration becomes extremely large if we use an imager with a large lens aperture. 
A combination of our 71 mm-aperture lens imager and 240 mm-diameter primary mirror, being restricted by our manufacturing capacity, results in a point-spread-function (PSF) size over one millimeter on the detector, which corresponds to 10 degrees in the sky.

We therefore did not completely follow the original two-mirror system scheme, and improved it to upgrade image quality at the cost of view near zenith and some amount of sensitivity.
In detail, we significantly flattened the vertex angle of the cone-shaped secondary mirror to reduce its tangential curvature.
Next, we introduced an aspherical shape similar to that of the secondary mirror in Figure \ref{opt_design1}, so that it also has a field-widening power to some extent, similar to a primary mirror. 
In fact, $O'$ projected by the section of this type of secondary surface does not strictly converge at one point, but its effect on the field deformation was found to be ignored.
Moreover, we adopted a smaller primary mirror hole to reduce the shadow area in the image center, and also to extend the focal depth.
Because the primary hole squeezes the imager's lens aperture, the decline in sensitivity is compensated by frame integration.

As a result of these improvements, we adopted the parameters listed for the ``Improved" model in Table \ref{sys_mirror}. 
The parameters determined following the original idea of Figure \ref{opt_design2} are also listed as a reference for the ``Original" model.
Our ``Improved" model gives a shadow circle with a radius of 11 degrees at the zenith. 
However, the area is not critical for us because the MAGNUM telescope rarely observes objects around the zenith because of the sparse distribution of celestial coordinates of our targets, and because of some operational restrictions of our telescope.

\begin{figure}[htbp]
\epsscale{0.75}
  \plotone{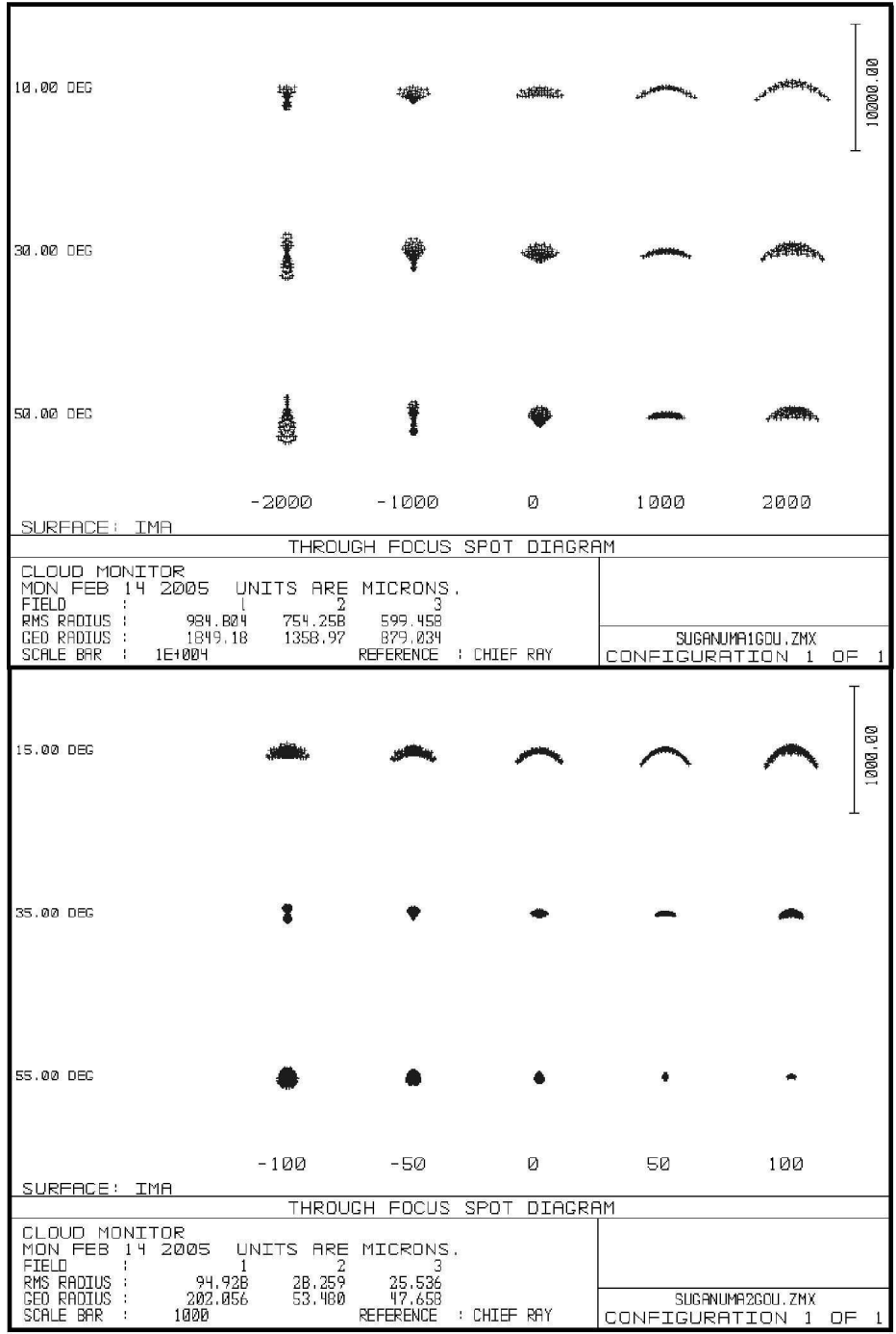}
  \caption{{\it Top: }Spot diagram through the focus for ``Original" model of the mirror system optimized to our camera, following Chahl \& Srinivasan (1997).
              {\it Bottom: }A similar diagram for ``Improved" model.
              Note that the unit of scale for the spots is micron, and the scale bar in the top panel is ten times larger than that in the bottom.  
              The pixel scale of the imager detector is 50 $\mu {\rm m}$.
             }
  \label{spt}
\end{figure}

Most remarkable is that the image resolution was dramatically increased in our ``Improved" model. 
Figure \ref{spt} shows the spot-diagrams of ray-traced images for both ``Original" and ``Improved" models.
The specifications of the two models are also tabulated in Table \ref{spe_mirror}.
The ``Improved" model decreases the size of the point-spread-functions (PSFs) by an order of magnitude or more.
There remain some astigmatism and field curvature in our ``Improved" model, but the PSF size is within a few pixels over almost the entire field of view.
\\

\subsubsection{Fabrication and Construction of Mirror System}

The surfaces of the mirrors were shaped by diamond turning on brass that is easily worked and goes well with gold coating. 
Using an ultra-precise, computerized, numerically controlled (CNC) turning machine, we obtained a surface roughness of about 20 nm in rms, which is fine enough as a mirror at a wavelength of $10\;\mu$m.
The surface was plated with solid gold, containing 5 \% Cobalt, with a thickness of 2 $\mu$m.
Finally, a physical vapor sapphire was deposited on the surface with a thickness of 0.2 $\mu$m for protection.
The surface reflectivity is about 95 \% and no serious degradation has been seen in an outdoor operation of five years.

For the first attempt to process the mirror surface, we tried an aluminum-based alloy plated by electroless nickel, and coated gold on the surface.
But we found that the gold coat degraded in a few months' exposure to air.
The pinholes in the nickel plating might make water erode the aluminum base rapidly.

On the central hole of the primary mirror was placed a chemical vapor-deposited (CVD) diamond plate with a thickness of 0.2 mm and transparency of about 80 \% at $\lambda \sim 10\mu {\rm m}$ with no coat.
A germanium plate processed with anti-reflection coating on both top and bottom and protective coating on the top could also work in some environments, and has been used in similar systems at several sites such as Mauna Kea in Hawaii and Atacama in Chile.
However, during our test operation in Tokyo, Japan, the upper surface of the germanium window degraded in a few months.
We guess that rain in Tokyo is very acidic, which might enhance degradation of the protective coating.
\\


\subsection{Data Acquisition System}
\label{acq_system}


The thermal imager, the Amber Sentinel Camera, outputs images with analog signals of NTSC standard as well as digital signals of 12-bit parallel channels at the rate of 30-frames per second.
We use digital output because an analog signal of 8-bit depth data loses the lower 4 bits of original signal that is much larger than the noise signal of one or two analog-to-digital units (ADUs), and is difficult to restore by frame integration afterwards.

For each data bit, together with synchronizing clocks for frame acquisition, the digital signals are single-ended transistor-transistor-logic (TTL) standard. 
We convert these signals to differential signals of RS-422 standard level by a hand-made digital electrical circuit with IC in order to transmit the signals to a PC about 15 m away from the outdoor system.
This is because the TTL signals are too delicate in a noisy environment to send more than several tens of centimeters at a high data rate.

The signals are acquired by a Linux PC using a digital frame-grabber board, PC-DIG produced by Coreco Inc.
It can grab digital data of 12 parallel channels at the rate of 4.6 mega-bytes per second for our imager, and its driver for Linux OS is supported by the company.

We integrate the images for 5 seconds, corresponding to 150 frames, to increase the signal-to-noise ratio.
Integration for more than 10 seconds is not favorable because the whole-sky images of clouds are often blurred by migrations of clouds.

For each acquisition of the whole-sky image, we take shutter images immediately before and after it.
In addition, the temperature of the blackbody reference plate is measured at the same time.

The sets of data are acquired about every one or two minutes while the elevation of the sun is below 15 degrees. 
We cease the operation almost entirely during the daytime, in case direct rays of sunlight degrade the imager detector.
\\


\section{ANALYSIS SOFTWARE}
\label{software}


\subsection{Data Reduction}
\label{reduction}

\begin{figure}[htbp]
\epsscale{0.95}
  \plotone{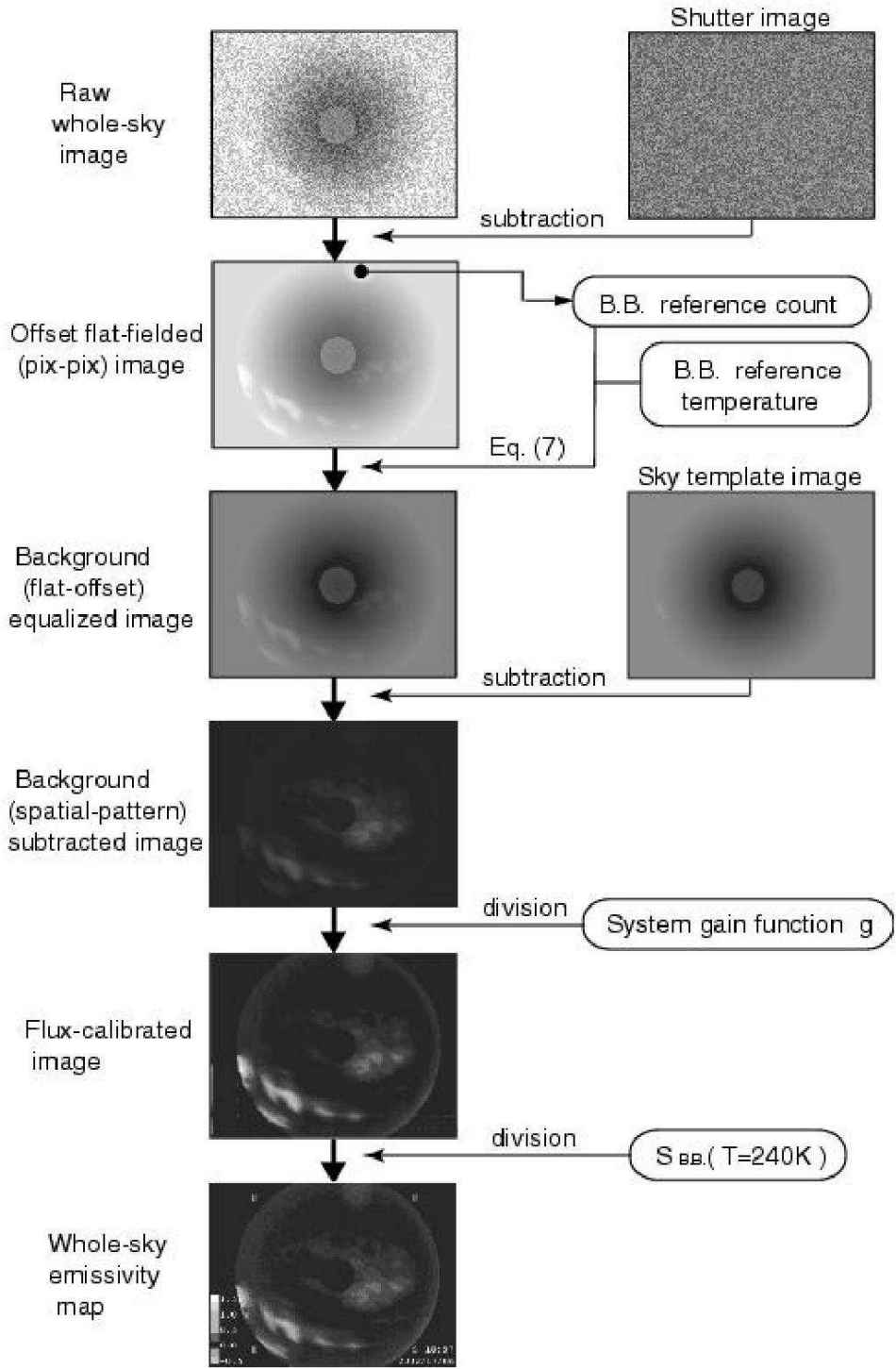}
  \caption{Block diagram of the reduction procedure from a raw whole-sky image into a whole-sky cloud emissivity map.}
  \label{kumored}
\end{figure}

Each whole-sky image is processed immediately after acquisition with the shutter images, the temperature of the blackbody reference, and some calibration data measured in advance.
Figure \ref{kumored} illustrates how we reduce a raw whole-sky image into the apparent emissivity map of the clouds.

First, offset flat fielding for the pix-to-pix pattern is done by subtracting the average of the two shutter images obtained before and after the incident exposure of the whole-sky image. 
The automatic offset flat fielding by the camera is convenient for snapshot images, but not complete for our frame-integrated whole-sky images.
We can do similar calibration with a higher signal-to-noise ratio using the frame-integrated shutter images.

Next, background signals from the optics and the interior of the imager are subtracted.
There are two types of components of background signal: flat-offset components and spatial-pattern components.
They differ from each other with varying internal and environmental temperatures, and have to be subtracted separately.

The flat-offset background component includes bias and dark current signals of the detector along with thermal radiation from the interior of the imager and the optics, which should be compensated for each whole-sky image.
Subtraction of the shutter image from the whole-sky image does not work well because the surface brightness of the shutter and the thermal background from the optics vary independently.
The flat-offset value of the background signal $C_{\rm off}$ on the whole-sky image is calculated from the temperature $T_{\rm ref}$ of the blackbody reference plate and its signal $C_{\rm ref}$ on the incident whole-sky image as
\begin{equation}
C_{\rm off} = C_{\rm ref} - \int B_{\lambda}(T_{\rm ref})\;d\lambda\;/\;g \;\;,
\label{gaincalib}
\end{equation}
where $B_{\lambda}(T_{\rm ref})$ is the Planck function for temperature $T_{\rm ref}$, and $g$ is the signal-to-surface brightness ratio measured beforehand. 
The units of $C_{\rm off}$ and $C_{\rm ref}$ are ADU.
We subtract the single value of $C_{\rm off}$ from those of all pixels in the incident whole-sky image.

There remains a spatial-pattern background component, which mainly originates in the baffle of the reflecting optics and the atmosphere. 
The pattern of this type of background is radially symmetrical and almost stable on clear nights.
We therefore prepare a template whole-sky image for a clear night beforehand.
The template image should be acquired when the sky is certain to be clear and reduced up to compensating the flat-offset background.
We subtract the template image from all whole-sky images. 

\begin{figure}[htbp]
  \epsscale{0.8}
  \plotone{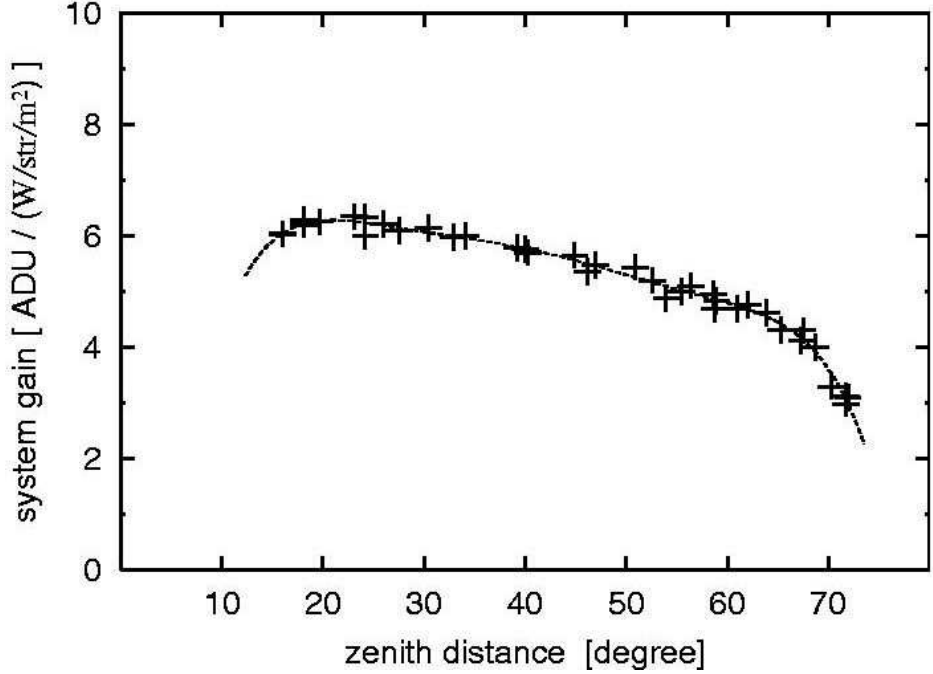}
  \caption{The signal-to-surface brightness ratio $g$ of MAGNUM infrared cloud monitor system, measured as a function of zenith angle.
    Dashed line, being fitted to the measured points, is used for $g$ in Figure \ref{kumored} and equation (\ref{gaincalib}). 
  }
  \label{gain}
\end{figure}

Now that the background signals are subtracted, we calibrate the signals in the whole-sky image into a surface brightness value using the signal-to-surface brightness ratio $g$.
One can determine the value of $g$ in laboratory by exposing the blackbody targets of different temperatures, or can determine it at the observatory site by exposing both clear sky and a black object of ambient temperature at once. 
Note that the value of $g$ is dependent on zenith angle, mainly because of vignetting on the aperture of the camera lens.
We should therefore measure $g$ for several zenith angles and using a function fitted to them.
Figure \ref{gain} shows the data of $g$ measured for our system at the MAGNUM observatory site.

Finally, the surface brightness $S$ in the image is converted to the apparent emissivity $\epsilon$ of the clouds at a 10 $\mu$m waveband, which is related to $S$ as
\begin{equation}
S=\epsilon \cdot \int B_{\lambda}(T_{\rm c})\; d\lambda \;\;\;,  
\label{bb}
\end{equation}
where $B_{\lambda}(T_{\rm c})$ is the Plank function for the cloud temperature $T_{\rm c}$.
According to the average annual air temperature of 296 K at sea level in Maui Island and lapse rate of $-6.5$ ${\rm K}/{\rm km}$ for standard atmosphere (COESA, U.S. Standard Atmosphere 1976), the expected ambient temperature at 10,000-m altitude above Haleakala Observatory should be about 240 K.
We therefore calculate $\epsilon$ assuming (hereafter fixing) the temperature of $T_{\rm c}=240$ K as representative of high-altitude clouds or cirrus.

\begin{figure}[htbp]
  \epsscale{0.9}
  \plotone{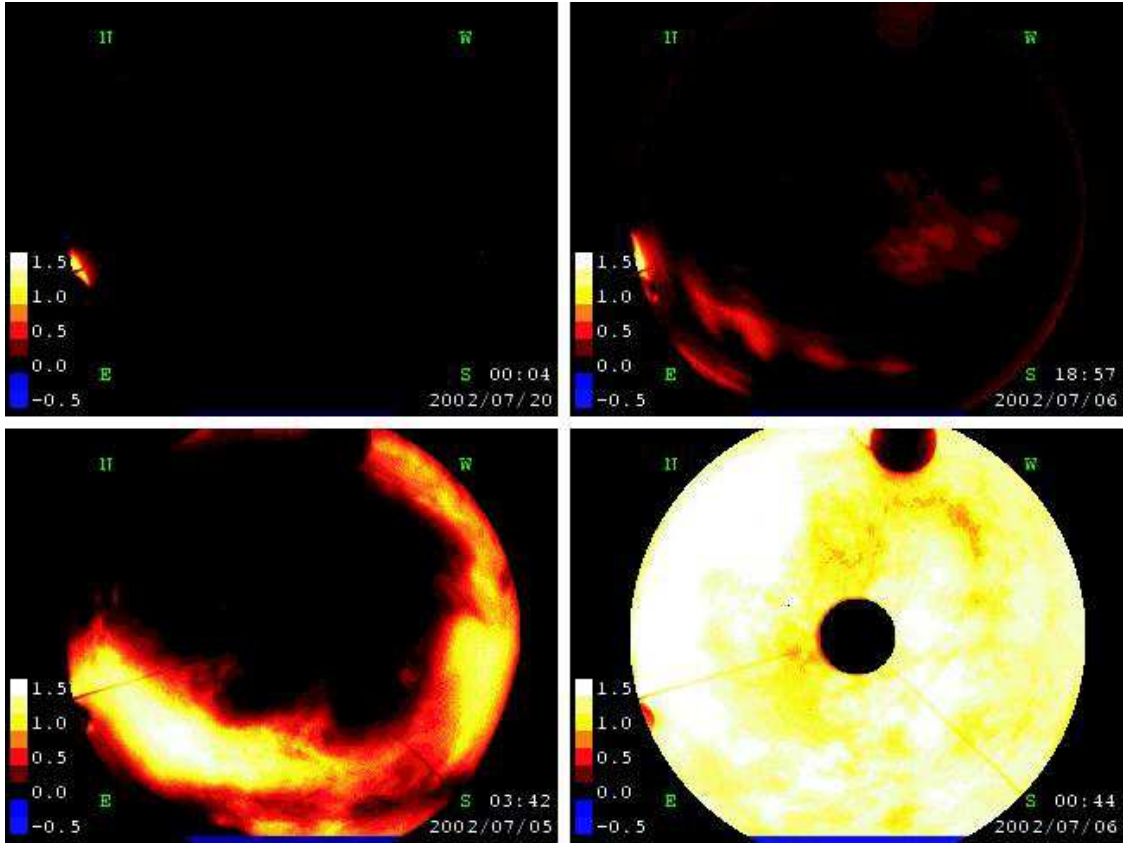}
  \caption{Whole-sky cloud emissivity maps acquired and processed by MAGNUM infrared cloud monitor under various sky conditions.
              The sky condition is clear at top left, thin at top right, partially cloudy at bottom left, and entirely cloudy at bottom right.
              Note that the emissivity is apparent one, which is defined as eq.\ref{bb}, being assumed $T_{\rm c}=240$ K.
              There are two shadow circles in each image: the small one at the image center is field vignetting by the hole of the primary mirror and the secondary mirror, and the other near the image edge is a blackbody reference plate.
}
  \label{cloudimage}
\end{figure}

Note that $\epsilon$ includes reflection efficiency of a cloud as well as absorption efficiency, and in $S$, there is a significant amount of reflected emission by the cloud which originates in the surface of the ground or sea.
This means we cannot convert the apparent emissivity $\epsilon$ simply to optical depth, which relates to actual absorption efficiency.
However, particularly for high-altitude clouds, it is reported that a large amount of emission still originates thermally in clouds themselves (Platt \& Stephens 1980). 
Figure \ref{cloudimage} shows the whole-sky cloud emissivity maps obtained and processed under various sky conditions.

It is very convenient for a remote watcher from Japan as well as at the Haleakala site to see the whole-sky cloud emissivity maps on the Internet.
However, our main objective in operating the MAGNUM observatory is automated observation in real-time consideration of weather conditions.
We therefore developed software to detect clouds from the whole-sky cloud emissivity maps and evaluate observational conditions from them ($\S$ \ref{acd} and $\S$ \ref{whole-sky}).
\\

\subsection{Automatic Cloud Detection}
\label{acd}

To determine whether clouds exist or not in a certain part of the sky, it is important to measure both the average and the standard deviation value of the emissivities in a small area in about that direction, rather than to refer to just one pixel value.
Here, two elements limit sensitivity: one is variation in zero emissivity level caused by residual thermal background signal, and the other is pix-to-pix noise.

The empirical value of the former for our system is about $\epsilon=0.25$, which is considerably large compared to those of thin clouds. 
This mainly comes from the residual pattern of the background radiation that is difficult to subtract completely from a single template whole-sky image.  
Also, humidity has some correlation with residual background.

The latter limit for the sensitivity is about $\epsilon=0.015$ as a noise equivalent signal of the image, which is much less than the former.
Thin clouds are easy to detect by their spatial fluctuations of emissivity rather than by emissivity values themselves.

We therefore divide a whole-sky cloud emissivity map into 90 sub-areas, each being 10 degrees in elevation and 20 degrees in azimuth.
For each sub-area, we categorize the cloud condition into several levels using the average emissivity $\overline{\epsilon}$ and the rms emissivity $\sigma(\epsilon)$ calculated for the area.  

\begin{figure}[htbp]
  \epsscale{0.6}
  \plotone{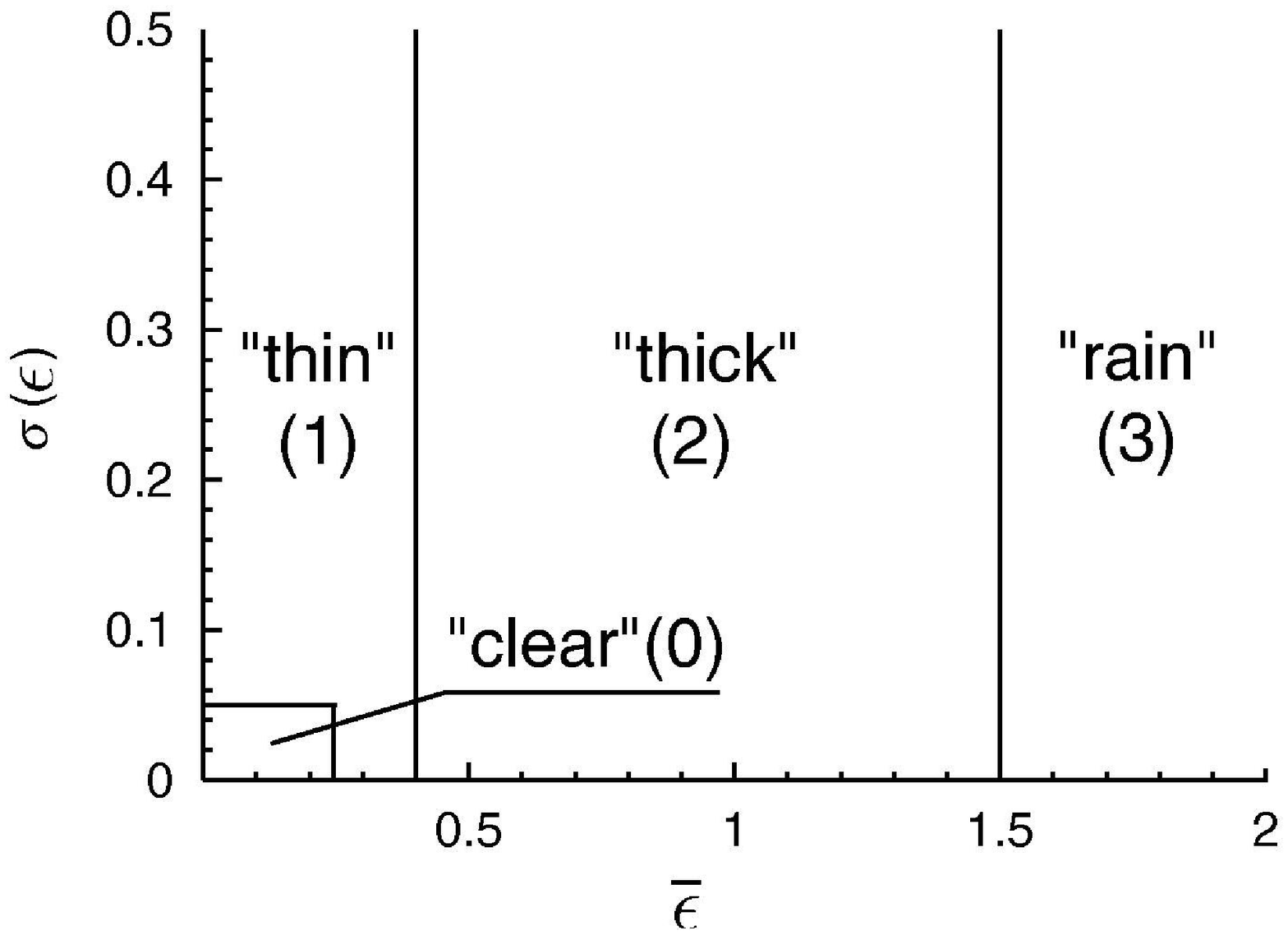}
  \caption{Classifications of cloud status for each sub-area of 20(Az)$\times$10(El) degrees from its average emissivity $\overline{\epsilon}$ and standard deviation $\sigma(\epsilon)$.
  A numerical value in parentheses in each zone indicates a level for calculating whole-sky cloud conditions (see \S\ref{whole-sky}). 
  }
  \label{pointsky}
\vspace{0.4in}
  \epsscale{0.65}
  \plotone{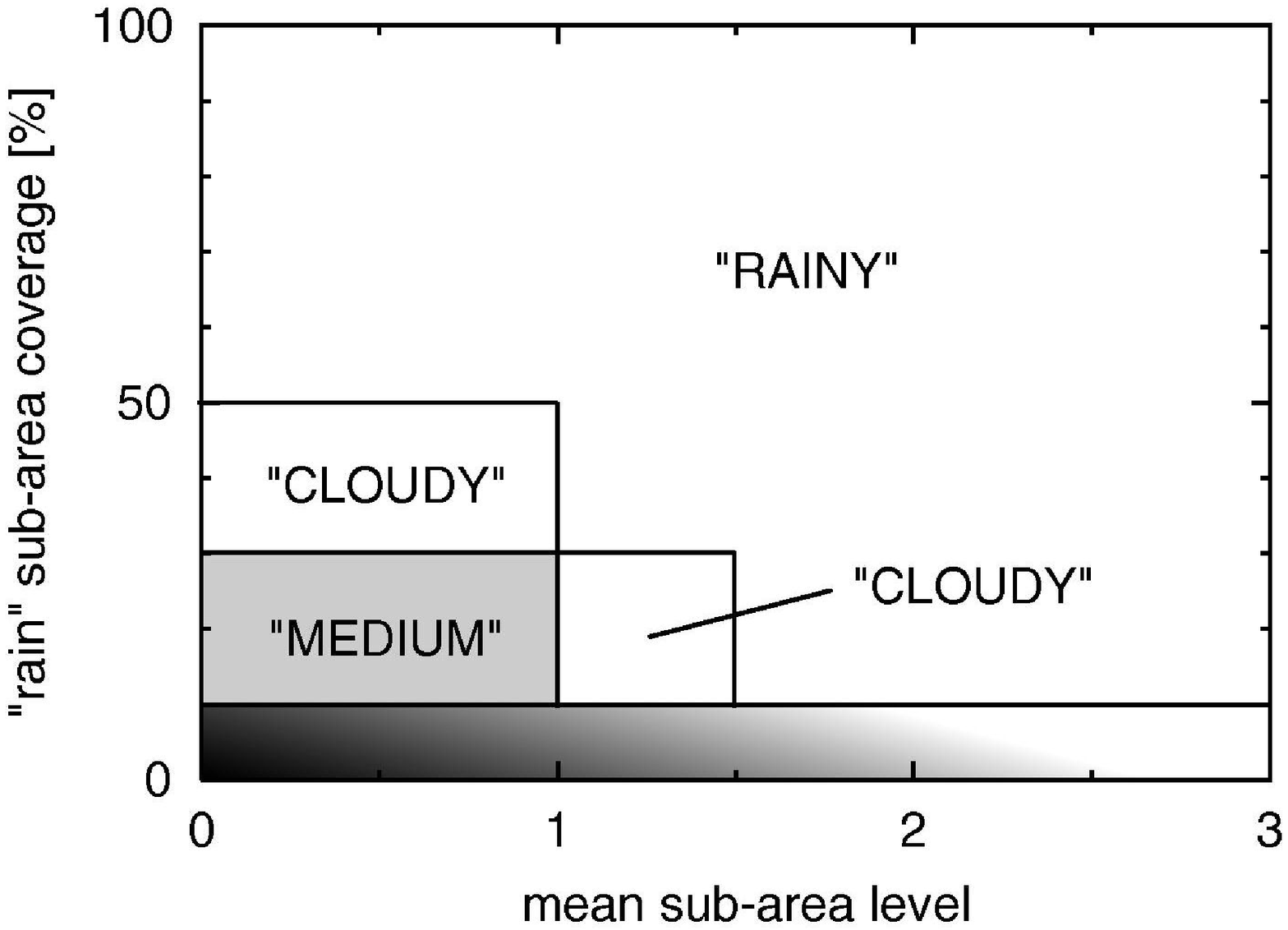}
  \caption{Classification of whole-sky cloud conditions from a ``rain" sub-area coverage and a mean sub-area level over all areas.
  The zone with gradation, whose ``rain" sub-area coverage is below 10 \%, is classified into several conditions on the other diagram (Fig.\ref{allsky_cloud}). 
  }
  \label{allsky_rain}
\end{figure}

Figure \ref{pointsky} shows the $\sigma$ versus $\epsilon$ diagram on which each sub-area can be evaluated.
A sub-area is evaluated as ``clear" only when both $\overline{\epsilon}<0.25$ and $\sigma(\epsilon)<0.05$, otherwise regarded as being covered by some clouds.
Except for ``clear", the sub-area is evaluated into ``thin", ``thick", or ``rain" when $\overline{\epsilon}<0.4$, $0.4\le\overline{\epsilon}<1.5$, and $\overline{\epsilon}\ge1.5$, respectively.
The condition of ``rain" means that the surface brightness is larger than that for blackbody of 300 K, and the mirror system is possibly wet due to rainfall or moisture, though the direct detection of rainfall should be done by rain sensors.
Each sub-area is given a level of 0 for ``clear", 1 for ``thin", 2 for ``thick", or 3 for ``rain" in order to calculate whole-sky cloud condition from statistics over all sub-areas (see next subsection).

The main cause preventing detection of even thinner clouds is the residual pattern of the thermal background on the whole-sky emissivity maps, which increases the rms value of emissivities even in a small sub-area.
This could be improved if a relation between the radial pattern and the temperature of the reflecting optics is contributed or the temperature of the reflecting optics is regulated.
More fundamentally, we are soon going to improve the design of the reflecting optics so that there would be no vignetting objects in the optical pass.
\\

\subsection{Classification of Whole-sky Cloud Condition}
\label{whole-sky}

To determine whether the sky allows observation or not and what type of observation is best to execute, we evaluate the whole-sky cloud condition using statistics over sub-area values calculated and labeled in $\S$ \ref{acd}.
We classify the whole-sky cloud condition into five types: ``CLEAR", ``THINorPARTIAL", ``MEDIUM", ``CLOUDY", and ``RAINY".

First, for safety, we strictly exclude conditions when there are many sub-areas of ``rain".
Figure \ref{allsky_rain} shows classifications of whole-sky cloud condition on the average over sub-area values vs. ``rain" sub-area coverage plane.
In so far as the ``rain" sub-area coverage is larger than 10 \%, the whole-sky condition is evaluated as ``MEDIUM", ``CLOUDY", or ``RAINY", and we do not start any type of observation. 

\begin{figure}[htbp]
  \epsscale{0.6}
  \plotone{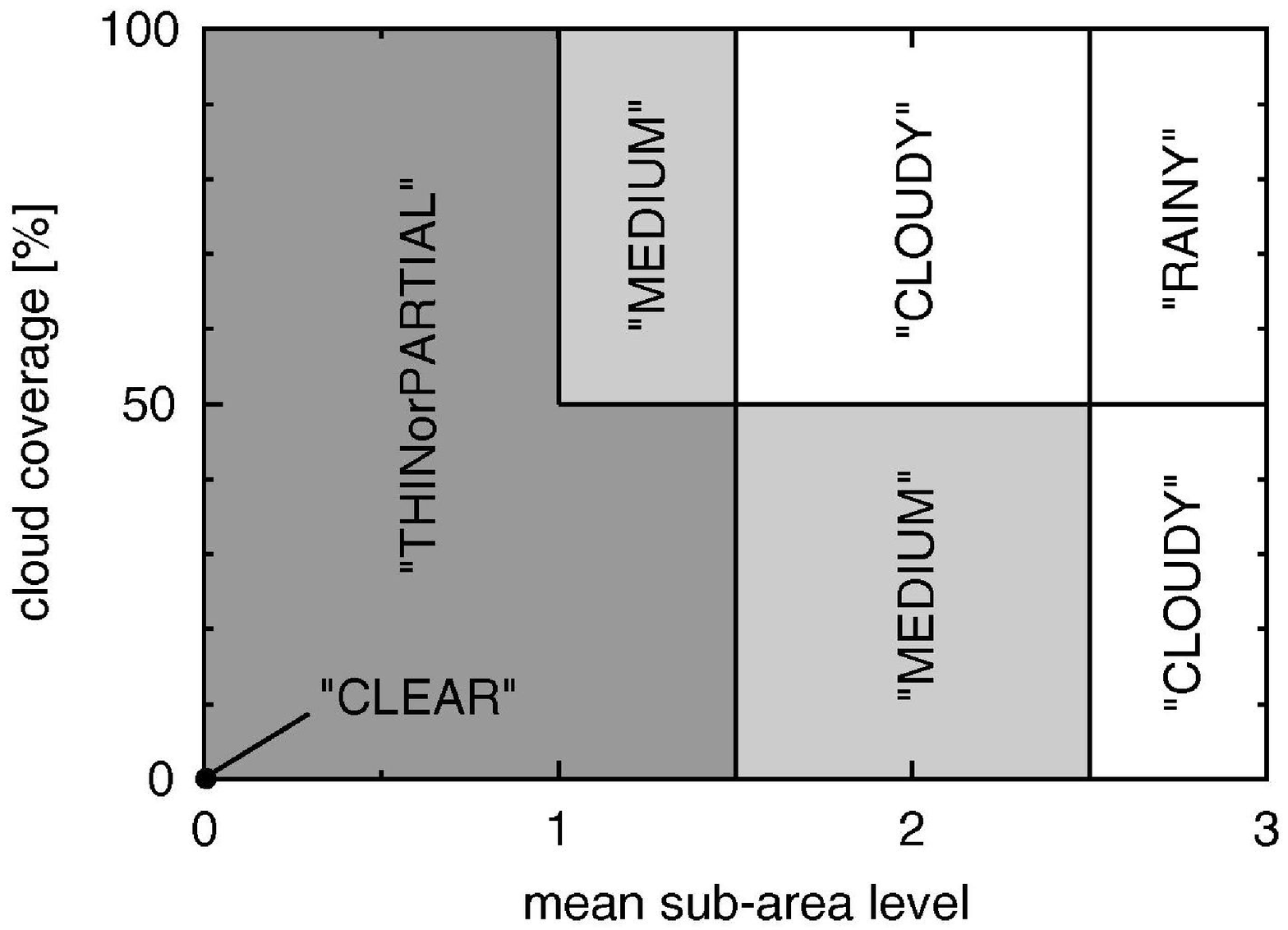}
  \caption{Classification of whole-sky cloud conditions from cloud coverage and a mean sub-area level over all areas.
  This classification is applied only when ``rain" sub-area coverage is below 10 \%.
  The cloud coverage includes both ``thin" and ``thick" sub-areas. 
  Only the origin of the diagram, where all the sub-areas are ``clear", is evaluated as ``CLEAR".
  }
  \label{allsky_cloud}
\vspace{0.2in}
  \epsscale{0.7}
  \plotone{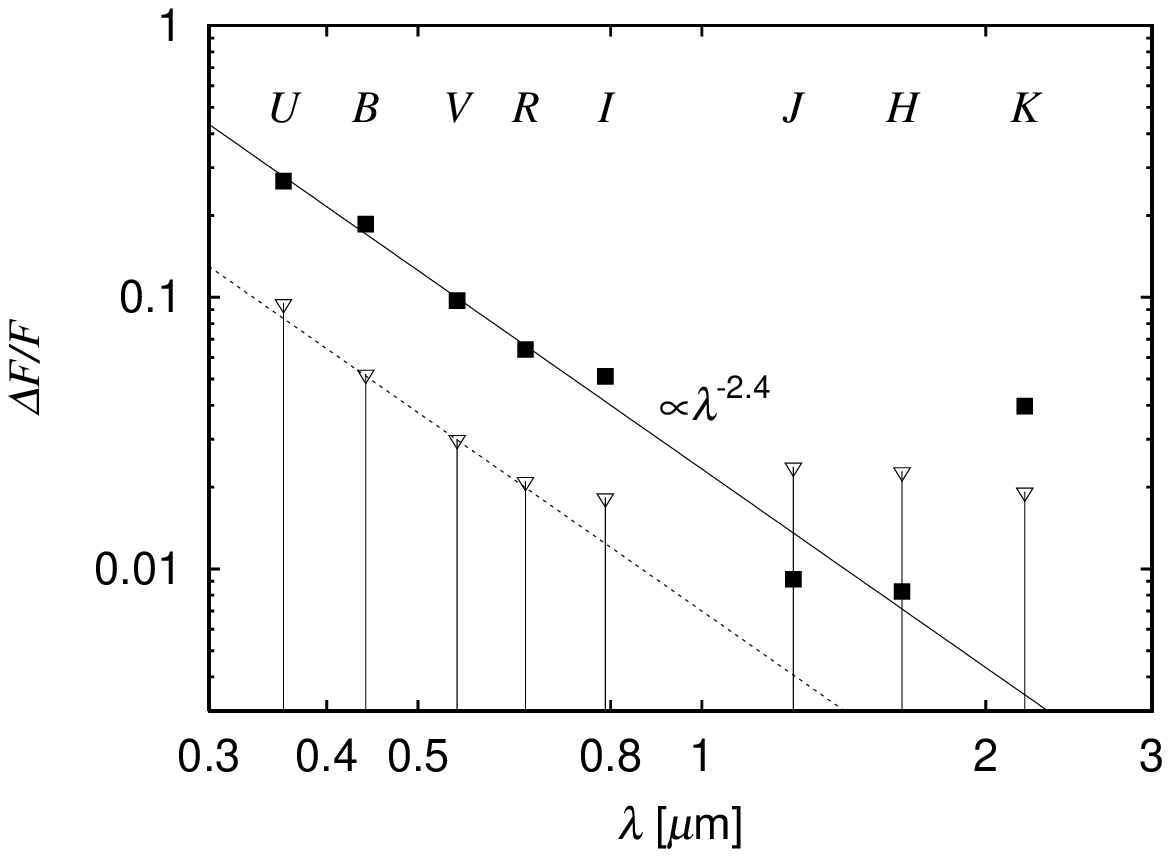}
  \caption{$\sigma_{\rm all}$ and $Q_{\rm atm}$ in Table \ref{sig_zerop}, plotted on wavelength vs. flux ratio plane.
           The inverted triangles with vertical lines are $\sigma_{\rm all}$, being correlated with $<$err$>$. 
           Filled squares are $Q_{\rm atm}$.
           The line fitted to the squares except for $K$-band shows wavelength dependence of $\lambda^{-2.4}$.  
           The dotted line is one downwarded by a factor of three from the fitted line. 
  }
  \label{std}
\end{figure}

Next, when the ``rain" sub-area coverage is less than 10 \%, we classify whole-sky cloud conditions into five types, as in Figure \ref{allsky_cloud}.
Here, cloud coverage includes both ``thin" and ``thick" sub-areas.
The whole-sky cloud condition is evaluated as ``CLEAR" only at the origin of Figure \ref{allsky_cloud}, which means all sub-areas are ``clear".
Except for ``CLEAR", whole-sky cloud condition is classified as ``THINorPARTIAL", ``MEDIUM", or ``CLOUDY", according to the cloud coverage and mean sub-area level.

The classifications above have mainly been working successfully, though they are empirical and somewhat inelegant.
\\


\section{PERFORMANCE}
\label{performance}


\subsection{Automated Observation with Infrared Cloud Monitor at MAGNUM Observatory}

Our cloud monitor was located at the Haleakala site and began to give whole-sky cloud emissivity maps when MAGNUM observatory started its telescope operation in August 2000.
Then, an automated monitoring observation of active galaxies with the cloud monitor was put into practical use in early 2001. 
After refinements of several months, we achieved fully automated astronomical observation for an entire night.
Now, we have continuous unmanned observation, except for maintenance every several months (Kobayashi et al. 2003; 2004) 

According to the whole-sky cloud condition evaluated by the cloud monitor, our automated observing system decides whether observation is possible or not.
When the whole-sky cloud condition is either ``CLEAR" or ``THINorPARTIAL", the observing system opens the dome slit and commands observation.
When the whole-sky cloud condition is ``CLOUDY" or ``RAINY", the observing system closes the dome slit, and carries out no observation. 
When the whole-sky cloud condition is ``MEDIUM", the observing system maintains the ongoing operation.

Moreover, according to the whole-sky cloud condition, we also determine what type of observation should be executed.
If the condition is ``CLEAR", which can be regarded as a photometric sky, all types of observations are possible.
If the condition is ``THINorPARTIAL", certain observations that are delicate under cloud extinction are restricted; observations such as standard star calibration, relative photometry between several separate fields, or imaging of faint objects are allowed only in ``CLEAR" conditions.
Instead, differential photometry between the bright objects in the same field of view is permitted in ``THINorPARTIAL" conditions because it is barely affected by extinction fluctuations.

The whole-sky cloud condition and status of the sub-area at which the telescope is pointing are recorded in the fits header of observed images.
The whole-sky emissivity maps are also archived so that we can check the quality of the observed astronomical data when we analyze them.

The cloud monitor has mainly been working stably until now, except for several months of trouble with the frame grabber board.
Regular maintenance includes wiping the dust on the mirror and shipping whole-sky images to Japan when we visit the site every several months.

In the following two sections ($\S$ \ref{defence_rain} and $\S$ \ref{det_phot}), we present and discuss performance of our cloud monitor, comparing it with some other weather sensors and photometric data.
\\

\subsection{Conservative Warning for Rainfalls}
\label{defence_rain}

The most primitive function required of our cloud monitor is to determine whether the sky allows observation or not.
When the sky becomes cloudy, the cloud monitor should close the dome slit before rain falls.
It also should stop meaningless and risky opening of the dome when the sky is still covered by thick clouds. 

Table \ref{sky_rain} shows the frequency distribution of various weather conditions over four years from two different weather-sensing systems including the whole-sky cloud monitor and the rain sensor.
Note that the rain sensor directly senses rain drops by means of changes in the resistivity of the electrical circuit, while the cloud monitor only inspects surface brightness of the sky image.
The percentages of rain-sensor output in the night are 85.8 \% for ``DRY" and 14.2 \% for ``RAIN".

This table indicates that a combination of ``CLOUDY" and ``RAINY" comprises 96 \% of ``RAIN", i.e., rain drops can be avoided by this high probability from such a combined cloud condition.
The remaining 4 \% probability corresponds to a situation in which the rain sensor catches rain drops while the output from the cloud monitor is ``CLEAR", ``THINorPARTIAL" or ``MEDIUM", and would decrease further if acquisition of whole-sky data were carried out more frequently, because the approach of moisture is sometimes very rapid.
The humidity sensor, however, usually helps to catch the moisture on its way up to the observatory.
\\

\subsection{Determination Whether the Sky is Photometric or Not}
\label{det_phot}

The next important function required of the cloud monitor is to determine whether the night is photometric or not.
The flux calibration of active galaxies using reference stars or standard stars in different telescope directions often fails if we are uncertain whether the sky is entirely clear.
Reliability of the whole-sky condition ``CLEAR" can be estimated from statistics of accumulated standard star flux data, because these have been observed quickly while the whole-sky is ``CLEAR".

Table \ref{sig_zerop} presents several statistical values of our standard star observations over two years while the instrumental throughput was relatively stable. 
Columns (1) and (2) are the waveband and effective wavelength, respectively. 
Column (3) is the number of observations. 
The standard deviation of the fluxes over all observations $\sigma_{\rm all}$ for each band is given in column (4), and the average over individual photometric errors $<$err$>$ is given in column (5). 
Nominal extinction value for unit airmass $Q_{\rm atm}$, measured by intensive observations of standard stars on a few nights, are shown in column (6).
The linear trend of flux decreasing with time during the period, derived from changes in telescope throughput, is corrected. 
Airmass correction for elevation in each observation is done with a constant value in the table.

The photometric errors $<$err$>$ are so small that they contribute little to $\sigma_{\rm all}$.
Therefore, the scatter $\sigma_{\rm all}$ mainly contains the day-to-day changes in extinction by the atmosphere or clouds.

Converting $\sigma_{\rm all}$ and $Q_{\rm atm}$ to flux ratio, we show $\Delta F/F$ against wavelength $\lambda$ in Figure \ref{std}.
The vertical bars with inverted triangles on top are $\sigma_{\rm all}$, being corrected for $<$err$>$. 
Filled squares are $Q_{\rm atm}$.
The solid line is linear fit to the filled squares except for the $K$-band which is particularly affected by water vapor.
The line shows wavelength-dependence of $\Delta F/F\propto\lambda^{-2.4}$, which is consistent with a typical trend of a mixture of Rayleigh-scattering by molecules and Mie-scattering by small aerosols in the atmosphere (Cox 2000).

A similar linear wavelength-dependence is seen in $\sigma_{\rm all}$ in the optical, and should be dominated by daily or seasonal variation in the extinction by the atmosphere.
On the other hand, $\sigma_{\rm all}$ at longer wavelengths beyond the $R$-band is near constant, regardless of wavelength-dependence of $Q_{\rm atm}$.
We consider that this flat component of $\sigma_{\rm all}$ could include variation in extinction by clouds missed by our cloud monitor, because the size of typical cloud particles is on the order of ten microns and their wavelength-dependence of extinction is white at a few microns or shorter.
Therefore, photometric errors caused by extinction of the clouds are restricted to within a few percent.
\\


\section{Trend of whole-sky cloud condition at Haleakala}
\label{weather_trends}


Bradley et al. (2006) overviewed meteorological characteristics at Haleakala with respect to many types of weather data, such as humidity, temperature, wind speed, and cloud coverage.
However, their analysis is based on compilation of various records with fairly large spatial and time resolution, including those taken by satellites. 
Our whole-sky cloud conditions are more straightforward and systematic, because our conditions are completely based on direct measurements of clouds that are projected onto the sky above the observatory.

Table \ref{log_allsky} shows the proportions of whole-sky cloud conditions averaged monthly between January 2001 and December 2005.
Several conditions in early 2001 are combined because of test operation of analysis software.
A total of 668,063 whole-sky images for 1,601 nights give the statistics in the table.

\begin{figure}[htbp]
  \epsscale{0.6}
  \plotone{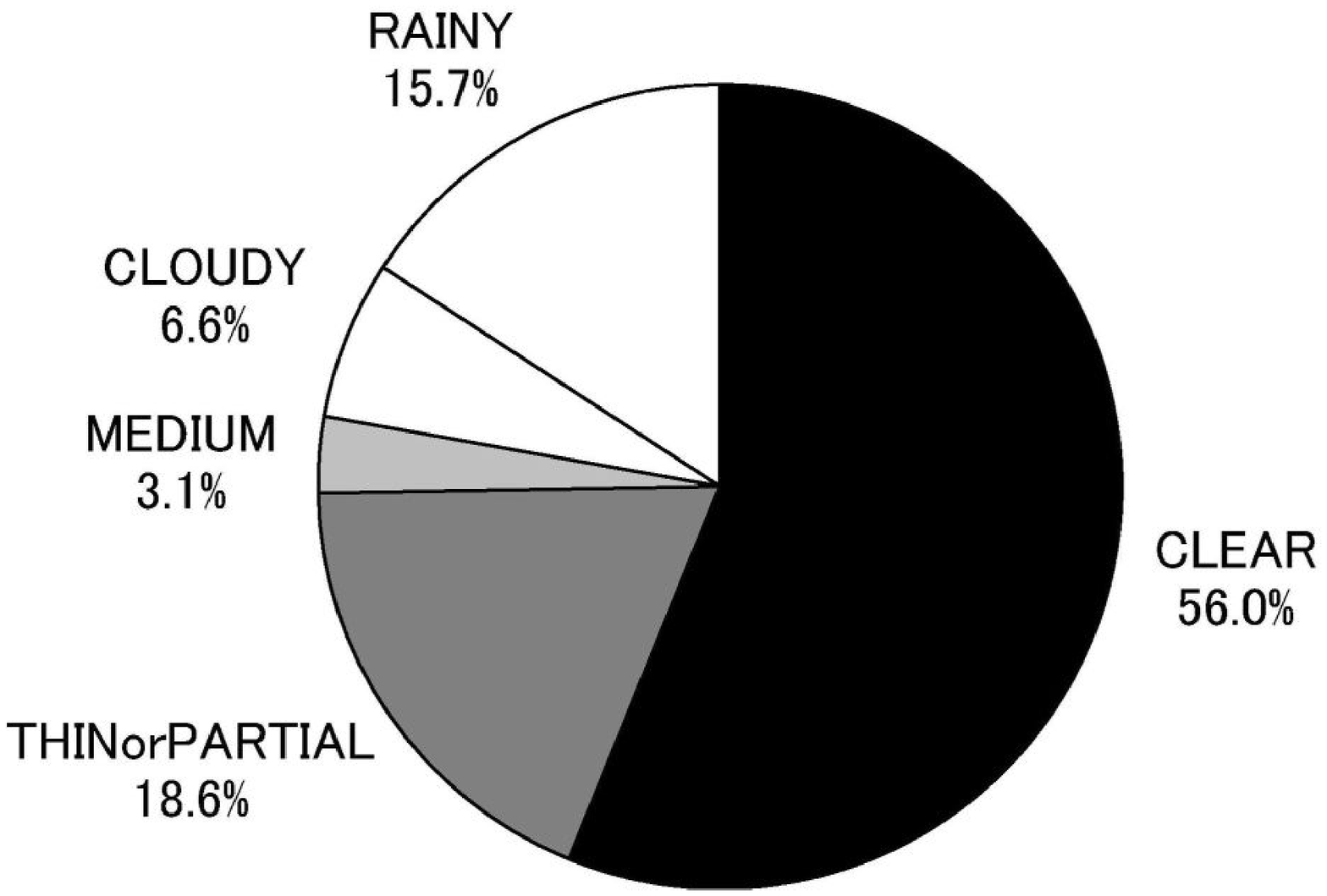}
  \caption{Distribution of whole-sky cloud conditions at night at Haleakala over five years between Jan 2001 and Dec 2005.
  Percentages shown are weighed by the number of nights per month.
  The percentage for combined conditions between January 2001 and July 2001 is divided into respective conditions, according to their average proportions after August 2001.
    }
  \label{weather_freq}
\vspace{0.4in}
  \epsscale{0.75}
  \plotone{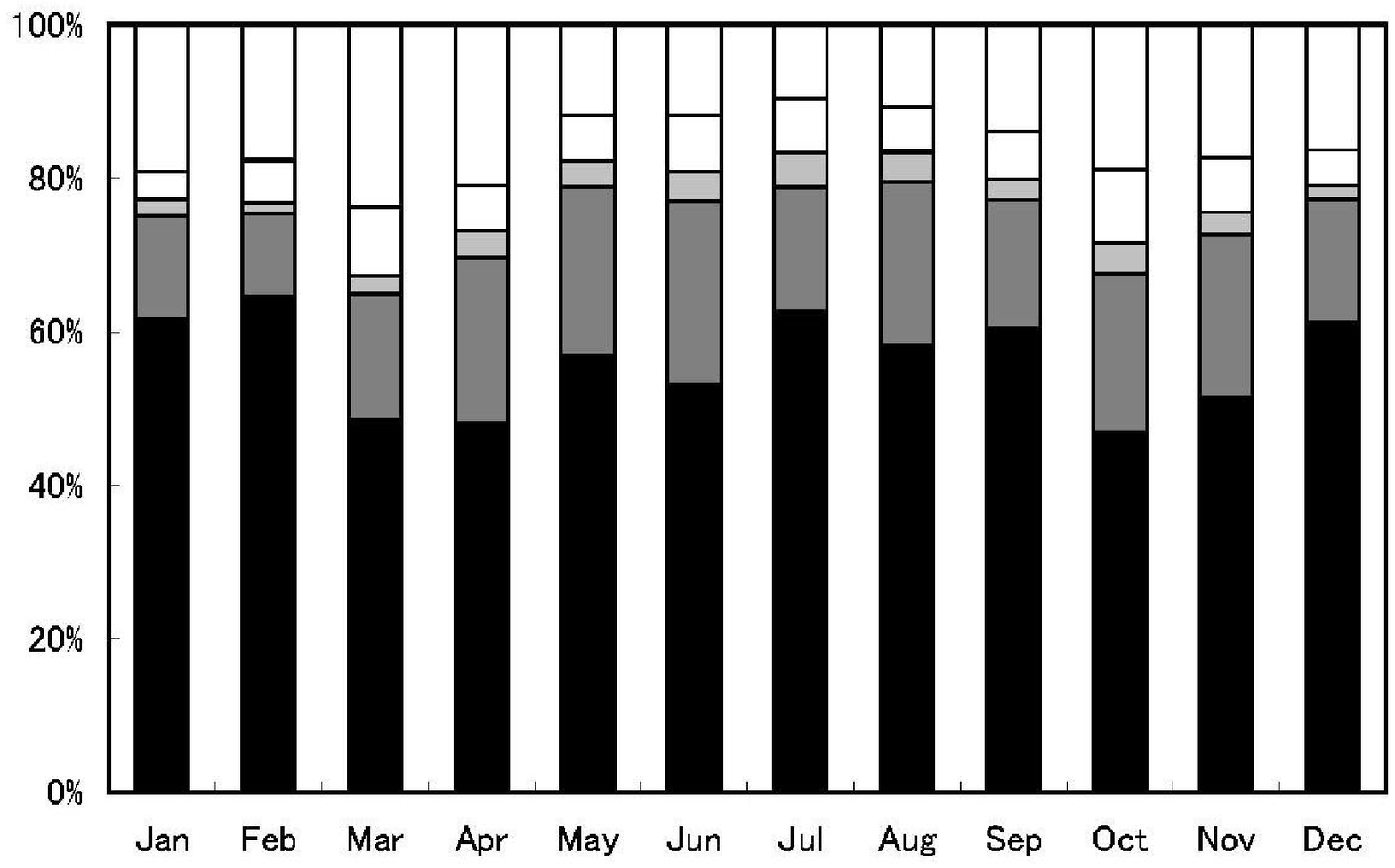}
  \caption{Mean monthly distribution of whole-sky cloud conditions at the Haleakala site over five years, from 2001 to 2005.
  From the bottom to top, ``CLEAR" (black), ``THINorPARTIAL" (gray), ``MEDIUM" (light gray), ``CLOUDY" (white), and ``RAINY" (white).
  The same corrections as in Figure \ref{weather_freq} were done for the data between 2001 January and 2001 July.
  }
  \label{monthly_weather}
\end{figure}

Figure \ref{weather_freq} presents the relative frequencies of the whole-sky cloud conditions combined over the data in Table \ref{log_allsky}.
The percentage of each condition is an average over the data weighted by the number of nights in which the data were obtained.
The monthly percentage for combined conditions between January 2001 and July 2001 is divided into respective conditions, according to their average proportions after August 2001.

It should be noted that despite ``CLEAR" and ``THINorPARTIAL" conditions, observations were sometimes impossible due to other weather warnings such as high humidity and strong wind.
Moreover, observations were not carried out when the wet sensor warned that a dome was not dried out after rainfall or moisture.
Concerning Haleakala, more than 50$-$60 \% of all night time is near photometric, and in about 75 \%, it  is feasible to execute particular observations.
Haleakala is therefore not worse than Mauna Kea where the observable sky rate is 60$-$80 \%\footnote{
http://www.naoj.org/Observing/Telescope/Image/seeing.html
} as one of the best locations for optical and near-infrared observations in the northern hemisphere, along with good access.

Next, Figure \ref{monthly_weather} shows the monthly average relative frequency of the whole-sky cloud condition.
Clear seasonal cycles over a year can be seen; There are high observable rates in summer and winter, and low rates in early spring and late autumn, in agreement with Bradley et al. (2006).
It has generally been said that there are a dry summer season and rainy winter season in Hawaii.
Our data, however, demonstrates that midwinter is not very bad at Haleakala, as far as sky condition is concerned.
\\


\section{Conclusions}


We developed an infrared cloud monitor weather system that has been most successfully supporting an unmanned robotic telescope.
It sees almost whole-sky in thermal infrared with no field deformation, sensitively detects thin high-altitude clouds, automatically evaluates sky conditions, and withstands outdoor environments for several months without maintenance.
Owing to this system, the MAGNUM observatory has been achieving unmanned automated observation at Haleakala for more than four years.
Its evaluation of the whole-sky cloud condition being photometric, observable, or non-observable seems to be mainly successful.
It also proves that for optical and near-infrared observations, Haleakala is a site comparable to Mauna Kea.
Copies of our cloud monitor are now used for many similar systems at sites all around the world, including the Atacama region in the northern part of Chile.

\acknowledgments

The development of our cloud monitor was supported by the Advanced Technology Center, National Astronomical Observatory of Japan. 
We are grateful to the Technical Center of Nagoya University for our use of their ultra-precise turning machine. 
We thank H. Takami and N. Takato for helpful discussion and advice on development. 
We also thank M. Doi, K. Motohara, and colleagues at the Haleakala Observatories for their help with facility maintenance. 
This research has been supported partly by the Grant-in-Aid of Scientific Research (10041110, 10304014, 11740120, 12640233, 14047206, 14253001, 14540223, 16740106, and 17104002) and COE Research (07CE2002) of the Ministry of Education, Science, Culture and Sports of Japan.




\begin{deluxetable}{ll}
\tablecolumns{2}
\tablewidth{0pc}
\tablecaption{Specifications of MAGNUM Infrared Cloud Monitor. \label{speicm}}
\tablehead{
}
\startdata
Wavelength & $8-14\mu {\rm m}$ \\
Optics & Thermal imager and two aspherical reflective\\ 
         & mirrors with Cassegrain-like alignment \\
Sensor & Micro-thermal bolometer array of 320 x 240pix \\
Field of View & Circular field of 11 - 70 degrees at zenith angle\\
Pixel Scale & 0.5 deg/pixel\\
Sampling rate & one-or two-minute interval \tablenotemark{a}\\ 
Sensitivity & $\epsilon\sim$ 0.015 for blackbody temperature of 240K\\
\enddata
\tablenotetext{a}{One image is integrated for 150 frames (total of 5 sec).}
\end{deluxetable}

\begin{deluxetable}{ll}
\tablecolumns{2}
\tablewidth{0pc}
\tablecaption{Specifications of the Infrared Imager (Amber Sentinel Camera). \label{sentinel}}
\tablehead{
}
\startdata
Detector & Uncooled micro bolometer array \\
Format & $320\times 240$ pixels \\
Wavelength & $8-14\mu $m \\
Lens & $f=50$ mm, {\it F}/0.7 $\rightarrow$ $\sim${\it F}/1.4 \tablenotemark{a}\\
Field of View & 18$^{\circ }\times 14^{\circ }$\\
NEDT\tablenotemark{b} & $<0.07$ K for blackbody temperature\\
                               & around 25$^{\circ }$C\\
Analog output & NTSC \\
Digital output & 12 bit parallel (TTL standard signals) \\
Remote control & RS232C interface\\
Frame rate & 30 Hz \\
\enddata
\tablenotetext{a}{The lens was stopped down to about {\it F}/1.4 by a mask in order to improve the image quality of the reflecting optics.}
\tablenotetext{b}{\underline{N}oise \underline{E}quivalent 
   \underline{D}ifferential \underline{T}emperature.}
\end{deluxetable}

\begin{deluxetable}{lcccccccccccc}
\rotate
\tablecolumns{13}
\tablewidth{0pc}
\tablecaption{Design parameters of the original and improved Cassegrain-like two mirror systems optimized for Amber Sentinel camera. \label{sys_mirror}}
\tablehead{
  && \multicolumn{6}{c}{Primary} & & \multicolumn{4}{c}{Secondary}\vspace{1mm}\\
  \cline{3-8}\cline{10-13}\\
Model && $O'(X, Z)$\tablenotemark{a} & $r'_{0}$ & $\gamma'_{0}$ & $\alpha$ & $\phi_{\rm out}$\tablenotemark{b} & $\phi_{\rm in}$\tablenotemark{c} & & $r_{0}$ & $\gamma_{0}$ & $\alpha$ & $\phi_{\rm out}$\tablenotemark{b}\vspace{0.5mm}\\
  && [mm] & [mm] & [degree] &  & [mm] & [mm] & & [mm] & [degree] &  & [mm] \vspace{0.5mm}
}
\startdata
Original .......  && (-52, 595)       & 530 & 85.0 & 11.0         & 240 & 80 && 300 & 85.0 & 1.0 & 80 \vspace{1mm}\\
Improved ..... && ($\;\,$-2, 488)  & 360 & 89.6 & $\;\;$4.4  & 240 & 38 && 340 & 89.6 & 2.3 & 100 \vspace{1mm}
\enddata
\tablenotetext{a}{the position of the virtual node $O'$ for primary in the coordinates $XOZ$}
\tablenotetext{b}{outer diameter}
\tablenotetext{c}{inner diameter, that is, the diameter of the center hole of the primary}
\end{deluxetable}

\begin{deluxetable}{lcccccc}
\tablecolumns{7}
\tablewidth{0pc}
\tablecaption{Specifications of the original and the improved mirror system optimized for Amber Sentinel camera. \label{spe_mirror}}
\tablehead{
Model && \multicolumn{2}{c}{Rms-size of PSF}  & & FOV\tablenotemark{a} & Focal\\
   &&    &    & &     &  ratio\\            
\cline{3-4}\\
   && [$\mu$m] & [pixels] && [degrees] &
}
\startdata
Original .....  &&  $600-1000$           & $12\;-20$ && $\;0-70$ & {\it F}/0.7$\;\;$\\
Improved ...  &&  $\;\;26-\;\;\;\,95$ & $\;\;0.5-2$  && $11-70$ & {\it F}/1.4
\enddata
\tablenotetext{a}{measured in zenith angle}
\end{deluxetable}

\begin{deluxetable}{crrrrrrrl}
\tablecolumns{9}
\tablewidth{0pc}
\tablecaption{Correlated classifications from two different weather systems \label{sky_rain}}
\tablehead{
Rain  &&  \multicolumn{5}{c}{Cloud Monitor} &&\\
Sensor &&&&&&\\
                \cline{3-7}\\
           && CLR & THN & MED & CDY & RNY && total$\;\;\;$
}
\startdata
DRY    && 55.8 & 18.0 & 2.5 & 5.9 & 3.5 && $\;\,$85.8\\
RAIN   && 0.1   & 0.4  & 0.2 & 0.7 & 12.9 && $\;\,$14.2\vspace{1mm}\\
\hline
total    && 55.9 & 18.4 & 2.7 & 6.6 & 16.4 && 100.0 \%\vspace{1mm}
\enddata
\tablecomments{CLR, THN, MED, CDY, and RNY mean the whole-sky cloud conditions, ``CLEAR", ``THINorPARTIAL", ``MEDIUM", ``CLOUDY", and ``RAINY", respectively.
                     Note that rain sensor directly senses rain drops or thick moisture, while the "RAINY" evaluated by cloud monitor means that there are high emissivity regions at 10$\mu$m waveband in the field of view.
                     }
\end{deluxetable}

\begin{deluxetable}{ccrccc}
\tablecolumns{6}
\tablewidth{0pc}
\tablecaption{Statistics of standard star observation \label{sig_zerop}}
\tablehead{
Band & Wavelength &  N-obs\tablenotemark{a}    & $\sigma_{\rm all}$\tablenotemark{b} & $<$err$>$\tablenotemark{c} & $Q_{\rm atm}$\tablenotemark{d}\\
       & [$\mu$m]   &                 &  [mag]     &  [mag]      &  [mag] \\
(1) & (2) & (3) & (4) & (5) & (6) 
}
\startdata
$U$ & 0.36 & 73  & 0.108 & 0.006 & 0.338\\
$B$ & 0.44 & 113 & 0.058 & 0.004 & 0.223\\
$V$ & 0.55 & 154 & 0.033 & 0.004 & 0.111\\
$R$ & 0.65 & 110 & 0.023 & 0.004 & 0.072\\
$I$ & 0.79 & 97   & 0.020 & 0.004 & 0.057\\
$J$ & 1.25 & 86   & 0.028 & 0.010 & 0.010\\
$H$ & 1.63 & 95  & 0.027 & 0.009 & 0.009\\
$K$ & 2.20 & 54  & 0.024 & 0.012 & 0.044\\
\enddata
\tablecomments{The observations are triggered only when the whole-sky condition calculated by the cloud monitor is "CLEAR".  
Airmass effects for elevation are corrected by $Q_{\rm atm}$.
Linear trends of decreasing flux during the two-year period of observations are also corrected.
}
\tablenotetext{a}{number of observations}
\tablenotetext{b}{standard deviations of flux over observations}
\tablenotetext{c}{average photometric error for each night}
\tablenotetext{d}{nominal extinction value per unit airmass measured at our observatory}
\end{deluxetable}

\begin{deluxetable}{clrrrrrrr}
\tablecolumns{9}
\tablewidth{0pc}
\tablecaption{Monthly average of the whole-sky cloud conditions during nights. \label{log_allsky}}
\tablehead{
Year &  Month & Nights & Ndata & CLR & THN & MED & CDY & RNY\\
\cline{5-9}
      &    &   &   & \multicolumn{5}{c}{[\%]}     
}
\startdata
2001 & Jan & 20 & 6981 & \multicolumn{3}{c}{90.4\tablenotemark{a}} & 3.6 & 6.0\\
       & Feb & 14 & 5121 & \multicolumn{3}{c}{68.3\tablenotemark{a}} & 5.7 & 26.0\\
       & Mar & 31 & 11097 & \multicolumn{2}{c}{83.0\tablenotemark{b}} & 2.3 & 2.6 & 12.1\\
       & Apr & 28 & 8946 & \multicolumn{2}{c}{62.5\tablenotemark{b}} & 5.5 & 6.3 & 25.7\\
       & May & 31 & 11732 & \multicolumn{2}{c}{93.6\tablenotemark{b}} & 1.9 & 0.8 & 3.7\\
       & Jun & 30 & 12351 & \multicolumn{2}{c}{65.9\tablenotemark{b}} & 9.5 & 10.3 & 14.3\\
       & Jul & 31 & 12328 & \multicolumn{2}{c}{62.8\tablenotemark{b}} & 12.3 & 10.4 & 14.5\\
       & Aug & 31 & 13051 & 67.0 & 25.3 & 2.4 & 2.7 & 2.6\\
       & Sep & 30 & 13210 & 55.1 & 25.1 & 3.1 & 8.8 & 7.9\\
       & Oct & 31 & 14540 & 47.1 & 22.7 & 2.4 & 8.8 & 19.0\\
       & Nov & 30 & 14424 & 46.9 & 22.9 & 2.1 & 9.4 & 18.7\\
       & Dec & 7 & 3427 & 0.0 & 76.3 & 3.6 & 10.5 & 9.6\\
2002 & Jan & 0 & 0 & ... & ... & ... & ... & ...\\
       & Feb & 0 & 0 & ... & ... & ... & ... & ...\\
       & Mar & 0 & 0 & ... & ... & ... & ... & ...\\
       & Apr & 0 & 0 & ... & ... & ... & ... & ...\\
       & May & 0 & 0 & ... & ... & ... & ... & ...\\
       & Jun & 10 & 3021 & 18.3 & 71.2 & 3.6 & 4.9 & 2.0\\
       & Jul & 31 & 11703 & 59.8 & 19.2 & 4.3 & 9.0 & 7.7\\
       & Aug & 31 & 9792 & 33.3 & 28.0 & 5.9 & 15.6 & 17.2\\
       & Sep & 30 & 9071 & 60.0 & 20.9 & 3.4 & 7.7 & 8.0\\
       & Oct & 31 & 9777 & 42.9 & 19.3 & 6.3 & 8.8 & 22.7\\
       & Nov & 30 & 12246 & 70.3 & 20.5 & 1.5 & 4.9 & 2.8\\
       & Dec & 31 & 14665 & 79.7 & 8.4 & 0.7 & 1.5 & 9.7\\
2003 & Jan & 31 & 14500 & 73.6 & 11.4 & 2.0 & 2.4 & 10.6\\
       & Feb & 28 & 12611 & 80.8 & 6.1 & 0.7 & 3.6 & 8.8\\
       & Mar & 31 & 13318 & 67.6 & 11.0 & 1.4 & 4.9 & 15.1\\
       & Apr & 30 & 12291 & 37.0 & 35.6 & 3.9 & 6.7 & 16.8\\
       & May & 27 & 10326 & 54.6 & 28.7 & 6.6 & 9.8 & 0.3\\
       & Jun & 30 & 11761 & 64.9 & 22.9 & 1.4 & 4.6 & 6.2\\
       & Jul & 31 & 12212 & 55.6 & 19.7 & 3.8 & 8.9 & 12.0\\
       & Aug & 31 & 12025 & 48.9 & 24.4 & 9.0 & 6.1 & 11.6\\
       & Sep & 29 & 11754 & 82.2 & 11.6 & 0.8 & 2.6 & 2.8\\
       & Oct & 31 & 13856 & 43.9 & 27.6 & 4.0 & 11.6 & 12.9\\
       & Nov & 30 & 13992 & 53.3 & 20.0 & 4.3 & 6.3 & 16.1\\
       & Dec & 31 & 14755 & 55.5 & 8.7 & 2.3 & 4.4 & 29.1\\
2004 & Jan & 31 & 14590 & 52.8 & 10.6 & 2.1 & 5.5 & 29.0\\
       & Feb & 29 & 13095 & 62.4 & 11.2 & 1.7 & 9.0 & 15.7\\
       & Mar & 31 & 13338 & 20.6 & 17.3 & 2.8 & 17.0 & 42.3\\
       & Apr & 30 & 12444 & 50.5 & 12.8 & 2.5 & 5.8 & 28.4\\
       & May & 31 & 12569 & 39.1 & 15.1 & 3.4 & 9.5 & 32.9\\
       & Jun & 30 & 11991 & 52.5 & 24.4 & 3.0 & 9.8 & 10.3\\
       & Jul & 31 & 12490 & 80.0 & 9.7 & 1.1 & 3.3 & 5.9\\
       & Aug & 31 & 12837 & 64.9 & 13.9 & 1.3 & 2.3 & 17.6\\
       & Sep & 30 & 12877 & 49.6 & 13.9 & 2.4 & 7.9 & 26.2\\
       & Oct & 31 & 13874 & 48.4 & 18.9 & 3.1 & 8.1 & 21.5\\
       & Nov & 30 & 13994 & 39.6 & 23.8 & 2.5 & 4.9 & 29.2\\
       & Dec & 31 & 14930 & 43.6 & 20.9 & 3.0 & 8.6 & 23.9\\
2005 & Jan & 31 & 14748 & 56.1 & 12.9 & 1.7 & 3.1 & 26.2\\
       & Feb & 28 & 12850 & 57.8 & 12.7 & 1.1 & 4.1 & 24.3\\
       & Mar & 31 & 13590 & 43.6 & 16.8 & 3.0 & 10.8 & 25.8\\
       & Apr & 30 & 12591 & 58.2 & 21.6 & 2.3 & 4.7 & 13.2\\
       & May & 31 & 12385 & 63.6 & 21.3 & 2.0 & 4.4 & 8.7\\
       & Jun & 30 & 11855 & 57.0 & 16.1 & 1.8 & 5.5 & 19.6\\
       & Jul & 31 & 12409 & 70.3 & 17.1 & 1.5 & 2.8 & 8.3\\
       & Aug & 31 & 12737 & 77.0 & 14.8 & 1.1 & 2.5 & 4.6\\
       & Sep & 30 & 12462 & 56.2 & 12.0 & 3.8 & 4.0 & 24.0\\
       & Oct & 31 & 13579 & 52.0 & 15.3 & 4.5 & 10.4 & 17.8\\
       & Nov & 30 & 14078 & 46.8 & 19.4 & 3.9 & 10.0 & 19.9\\
       & Dec & 31 & 14866 & 79.5 & 12.9 & 0.7 & 3.1 & 3.8\\
\enddata
\tablecomments{CLR, THN, MED, CDY, and RNY mean the whole-sky cloud conditions, ``CLEAR", ``THINorPARTIAL", ``MEDIUM", ``CLOUDY", and ``RAINY", respectively.
                     No data other than analog vision were obtained between January 2002 and May 2002.}
\tablenotetext{a}{combined with CLR, THN, and MED.}
\tablenotetext{b}{combined with CLR and THN.}
\end{deluxetable}

\end{document}